\RequirePackage{fix-cm}
\documentclass[twocolumn]{svjour3}          
\smartqed  
\makeatletter
\def\cl@chapter{\@elt {theorem}}
\makeatother
\usepackage{graphicx}
\usepackage{mathptmx}      
\usepackage{float}
\usepackage{amsmath}
\usepackage{microtype}
\usepackage{xfrac}
\usepackage{textcomp}
\usepackage{multirow}
\usepackage{booktabs}
\usepackage{siunitx}
\usepackage{rotating}
\usepackage{gensymb}
\usepackage{dirtytalk}
\usepackage{import}
\usepackage{tikz}
\usepackage{pgf}
\usepackage{pgfplots}
\usepackage[compatibility=false]{caption}
\usepackage[hypcap,labelformat=simple]{subcaption}
\usepackage[]{natbib}
\usepackage{enumerate}
\journalname{Int J Comput Vis}
\usepackage[
pdfpagelabels,hypertexnames=true,
plainpages=false,
naturalnames=false,]{hyperref}
\usepackage{cleveref}

\usepackage{color}
\definecolor{darkblue}{rgb}{0,0.1,0.5}
\hypersetup{colorlinks,
	linkcolor=red,
	anchorcolor=blue,
	citecolor=blue,
	urlcolor=blue}
\newcommand{\RM}[1]{\MakeUppercase{\romannumeral #1{.}}}
\newcommand*{\figWidthOneCols}{.8}

\newcommand*{\tabWidthTwoCols}{1}

\Crefname{equation}{Eq.}{Eqs.}
\Crefname{figure}{Fig.}{Figs.}
\crefname{figure}{}{Figures}
\Crefname{tabular}{Tab.}{Tabs.}
\usepackage{url}
\makeatletter
\g@addto@macro{\UrlBreaks}{\UrlOrds}
\makeatother
\begin{document}

\title{Baseline and Triangulation Geometry in a Standard Plenoptic Camera\thanks{This research was supported in part by the EU \if under the ICT program\fi as Project \textit{3D~VIVANT} under EU-FP7 ICT-2010-248420.}}


\author{\mbox{Christopher Hahne {$\enspace\cdot\enspace$} Amar Aggoun {$\enspace\cdot\enspace$} Vladan Velisavljevic {$\enspace\cdot\enspace$} Susanne Fiebig {$\enspace\cdot\enspace$} Matthias Pesch}}
\authorrunning{Christopher Hahne et al.} 

\institute{Christopher Hahne$^{1}$ \at
              \email{info@christopherhahne.de}           
           \and
           Amar Aggoun$^{1}$ \at
           \email{amar.aggoun@beds.ac.uk}      
           \and
           Vladan Velisavljevic$^{1}$ \at
           \email{vladan.velisavljevic@beds.ac.uk}           
           \and
           Susanne Fiebig$^{2}$ \at
           \email{sfiebig@arri.com}           
           \and
           Matthias Pesch$^{2}$ \at
           \email{mpesch@arri.com}           
           \and
            $^{1}$ University of Bedfordshire, Luton, United Kingdom \\
            $^{2}$ ARRI Cine Technik GmbH \& Co. KG, Munich, Germany \\
}

\date{Received: date / Accepted: date}

\maketitle

\begin{abstract}
In this paper, we demonstrate light field triangulation to determine depth distances and baselines in a plenoptic camera. %
%
Advances in micro lenses and image sensors have enabled \if so-called\fi plenoptic cameras to capture a scene from different viewpoints with sufficient spatial resolution. 
While object distances can be inferred from disparities in a stereo viewpoint pair using triangulation, this concept remains ambiguous when applied in the case of plenoptic cameras. %
We present a geometrical light field model allowing the triangulation to be applied to a plenoptic camera in order to predict object distances or specify baselines as desired. 
It is shown that distance estimates from our novel method match those of real objects placed in front of the camera. Additional benchmark tests with an optical design software further validate the model's accuracy with deviations of less than $\pm0.33~\%$ for several main lens types and focus settings. %
A variety of applications in the automotive and robotics field can benefit from this estimation model. 
\end{abstract}
\section{Introduction}
\label{intro}
%
%
Computer vision has been striving to recreate our human visual perception. %
Wheatstone's fundamental observations \citep{Wheatstone:1838} state that a set of solely two adjacent cameras facilitates imitating a human's binocular vision. Using these two images in conjunction with a \if matching state-of-the-art\fi stereo display technique, e.g. stereoscopic glasses~\citep{HUANG:2015}, allows for the reproduction of depth as perceived by humans. With regard to the location in object space, however, this stereo vision system concedes much more freedom than the human's perception as the distance between cameras, called \textit{baseline}, may vary. Hence, the flexibility in camera stereoscopy makes it possible to adapt to particular depth scenarios. 
For example, triangulation is used in \if the so-called \fi stellar parallax to measure the distance to stars~\citep{Hirshfield}. What applies to a macroscopic universe, may also be useful for a \if plenoptic\fi microscope. \par%
However, miniaturising multiple stereo setups to the level as required by microscopes poses a problem to hardware fabrication since lens diameters restrict baseline gaps between cameras. As an alternative, a \textit{Micro Lens Array}~(MLA) may be placed in front of an image sensor of an otherwise conventional microscope~\citep{Levoy:06,Broxton:13}, which is generally known as a light field camera. An obvious attempt to regard the micro lens pitch as the baseline proves to be impractical as optical parameters of the objective lens affect a light field's geometry~\citep{Hahne:14:IEEE,Hahne:14:OPEX}. %
\par
%
The light field camera, also known as plenoptic camera, was adopted to the field of computer vision ever since \citeauthor*{AW}~(\citeyear{AW}) published an article, which coined the term \textit{plenoptic} deduced from Latin and Greek meaning \say{full view}. 
The authors were the first to computationally generate a depth map by solving the stereo correspondence problem based on footage from a plenoptic camera and concluded that its baseline is confined to the main lens' aperture size. Although \citeauthor*{AW} could not provide methods to acquire quantitative baseline measures, the authors predicted the baseline to be relatively small. When \citeauthor*{LEVHAN}~(\citeyear{LEVHAN}) proposed a concise \mbox{4-D} light field notation, each ray in the light field could be represented by merely four coordinates $(u,v,s,t)$ obtained from the rays' intersection at two two-dimensional {(2-D)} planes placed behind one another. In respect of a plenoptic camera, these sampling planes may be represented by MLA and image sensor. In case of a plenoptic camera, maximum directional light field resolution is captured when focusing micro lenses to infinity~\citep{NG}, which is accomplished by placing the MLA stationary one focal length in front of the sensor. This plenoptic camera type has been made commercially available by \citeauthor*{LYTRO}~(\citeyear{LYTRO}) and is capable of synthetically focusing images~\citep{NGLEV,Fiss:14:ICCP,Hahne:16:OPEX}. %
By shifting the sensor away from the MLA focal plane, research has shown that the spatial and directional resolution can be traded off, which involves different image synthesis approaches~\citep{LUMSFULL,GEOSPATIO}. To distinguish between these optical setups, Lytro's camera was later named \textit{Standard Plenoptic Camera}~(SPC) in a publication by \citeauthor*{RAYTRIX}~(\citeyear{RAYTRIX}), who devised a more complex MLA that features different micro lens types. 
The spatio-angular trade-off in a plenoptic camera is determined by diameter, focal length, image position and packing of the micro lenses, just as the sensor pixel pitch, which thus makes it part of the optical hardware design.
\par
Over the years, several studies have provided different methods to acquire disparity maps from an SPC~\citep{heber:eccv14,bok:eccv14,jeon:cvpr15,Tao:IEEE:16}. To the best of our knowledge, researchers have not dealt with the estimation of an object's distance using triangulation on the basis of disparity maps obtained from a light field camera. One reason might have been that baselines are required, which are not obvious in the case of plenoptic cameras as the optics involved is more complex than with conventional stereoscopy. %
Attempts to estimate a plenoptic camera's baseline were initially addressed in publications by our research group~\citep{Hahne:14:IEEE,Hahne:14:OPEX}, which provided validation through simulation only. Besides, main lens pupil positions have been ignored in this work, yielding large deviations when estimating the distance to refocused image planes obtained from an SPC~\citep{Hahne:16:OPEX}. It is thus expected that our previous triangulation scheme \citep{Hahne:14:IEEE,Hahne:14:OPEX} entails errors in the experimentation which is subject to investigation. A more recent study by \cite{jeon:cvpr15} has also proposed a baseline estimation method without giving details on the optical groundwork and lacking validation activities. 
\par
\begin{figure}[H]
	\centering
	\includegraphics[width=.75\linewidth]{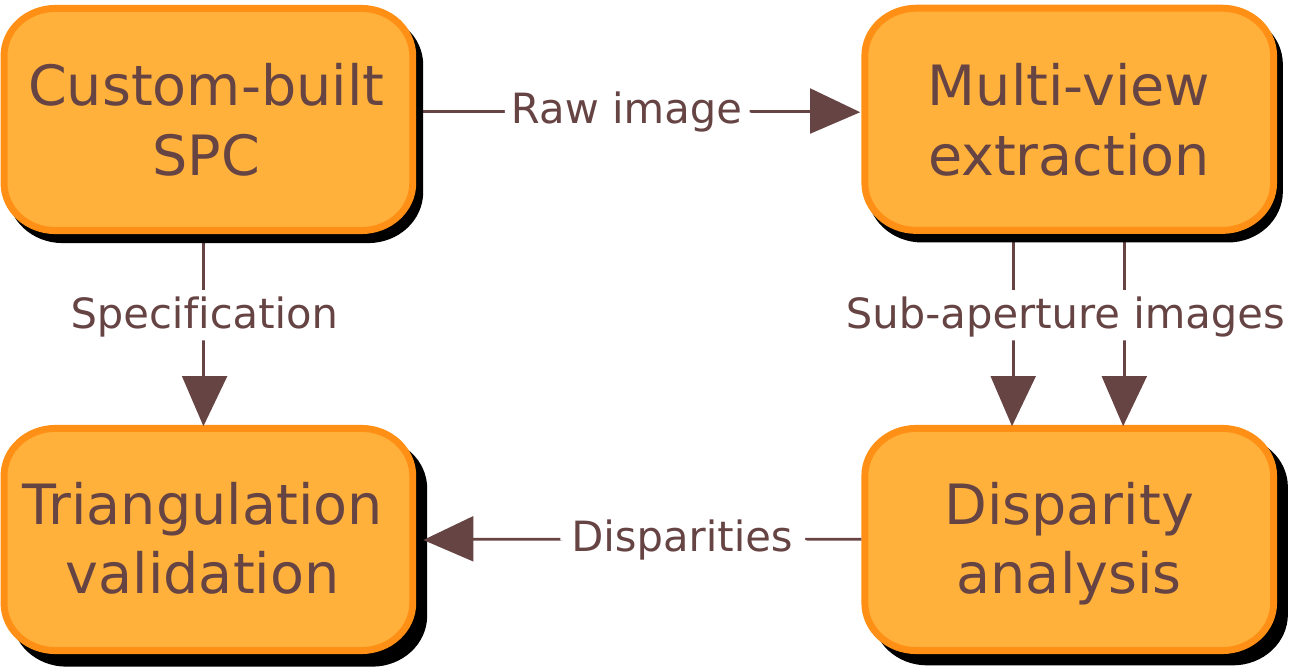}
	\caption{Block diagram for experimental model validation. \label{fig:top-level}}
\end{figure}

In this paper, we propose a refined optics-geometrical model for light field triangulation and estimate object distances captured by an SPC. Our plenoptic model is the first to pinpoint virtual cameras along the entrance pupil of the objective lens. Verification is accomplished through real images from a custom-built SPC and a ray tracing simulator~\citep{Zemax} for a quantitative deviation assessment. 
A top-level overview of the processing pipeline for experimental validation is given in Fig.~\ref{fig:top-level}.
By doing so, we obtain much more accurate baseline and object distance results than by our previous method~\citep{Hahne:14:IEEE} and \cite{jeon:cvpr15}. %
The proposed concept will prove to be valuable in fields where stereo vision is traditionally used. 
\par
%
This paper has been organised in the following way. Section~\ref{sec:1} briefly reviews the binocular vision concept by means of the geometry in order to recall stereo triangulation. This is followed by a step-wise development of an SPC ray model in Section~\ref{sec:2} where the extraction of viewpoints images from a raw SPC capture is also demonstrated. Experimental work is presented in Section~\ref{sec:3}, which aims to assess claims made in Section~\ref{sec:2} by measuring baseline and tilt angle from a disparity map analysis and a ray tracing simulation~\citep{Zemax}. Results are summarised and discussed in Section~\ref{sec:4}. %
\section{Stereoscopic Triangulation}
\label{sec:1}
\subsection{Coplanar Stereo Cameras}
The SPC can be seen as a complex derivative of a stereo vision system. The stereo triangulation concept is presented hereafter to serve as a groundwork. \par%
%
\begin{figure}[ht]
	\centering
	\includegraphics[height=3.5cm]{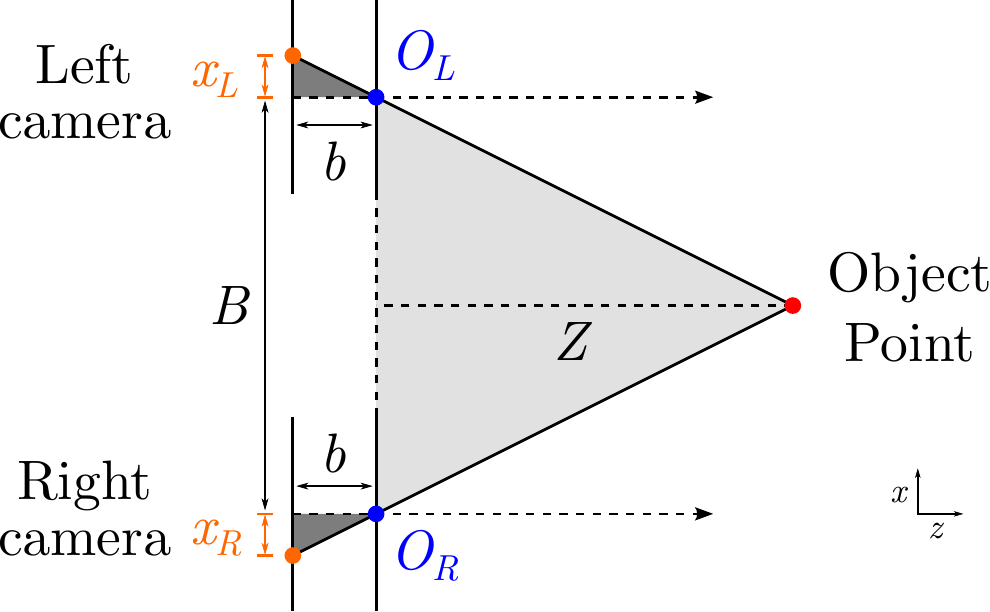}
	\caption[Stereo triangulation with parallel cameras]{Stereo triangulation scheme with parallel cameras where a point is projected through the optical centres $O_L$, $O_R$ yielding two image points (orange) in each camera. The relative displacement of these points returns the horizontal disparity $\Delta x = x_R - x_L$. The baseline $B$, object distance $Z$ and image distance $b$ affect the measured disparity.}
	\label{fig:stereoscopy}
\end{figure}
Figure~\ref{fig:stereoscopy} illustrates a stereoscopic camera setup where sensors are coplanar. The depicted setup may be parameterised by the spacing of the cameras' axes, denoted as $B$ for baseline, the cameras' image distance $b$ and the optical centres $O_L$, $O_R$ for each camera, respectively. As seen in the diagram, an object point is projected onto both camera sensors indicated by orange dots. With regard to corresponding image centres, the position of the image point in the left camera clearly differs from that in the right. This phenomenon is known as \textit{parallax} and results in a relative displacement of respective image points from different viewpoints. To measure this displacement, the horizontal disparity $\Delta x$ is introduced given by $\Delta x = x_R - x_L$, where $x_R$ and $x_L$ denote horizontal distances from each projected image point to the optical image centre. Nowadays, image detectors are composed of discrete photosensitive cells making it possible to locate and measure $\Delta x$. The disparity computation is a well studied task~\citep{Marr:1976:CCS:889236, YANG:93,Bobick:1999:LOS:335178.335180} and is often referred to as solving the correspondence problem. Algorithmic solutions to this are applied to a set of points in the image rather than a single one and thus yield a map of $\Delta x$ values, which indicate the depth of a captured scene. \par%
An object point's depth distance $Z$ can be directly fetched from parameters in Fig.~\ref{fig:stereoscopy}. As highlighted with a dark tone of grey, $\Delta x$ may represent the base of any acute scalene triangle with $b$ as its height. Another triangle spanned by the base $B$ and height $Z$ is a scaled version of it and shown in light grey. This relationship relies on the method of similar triangles and can be written as an equality of ratios
\begin{align}
	\centering
	\frac{Z}{B} = \frac{b}{\Delta x} \label{eq:triangleRatio} \, .
\end{align}
To infer the depth distance $Z$, Eq.~(\ref{eq:triangleRatio}) may be rearranged to
\begin{align}
	Z = \frac{b \times B}{\Delta x} \, . \label{eq:classicTria}
\end{align}
As seen by these equations, it is feasible to retrieve information about the depth location $Z$. Likewise, if $\Delta x$ is constant, it may be obvious that by decreasing the baseline $B$, the object distance $Z$ shrinks. Given a case where the depth range is located at a far distance, it is thus recommended to aim for a large baseline. Note that this relationship and corresponding mathematical statements only hold for cases where optical axes of $O_L, O_R$ are aligned in parallel. %
%
\subsection{Tilted Stereo Cameras}
Reasonable scenarios exist in which a camera's optical axis is tilted with respect to the other. 
In such a case, the principle of similar triangles does not apply in the same manner as in Eq.~(\ref{eq:triangleRatio}).
\begin{figure}[H]
	\centering
	\includegraphics[height=3.5cm]{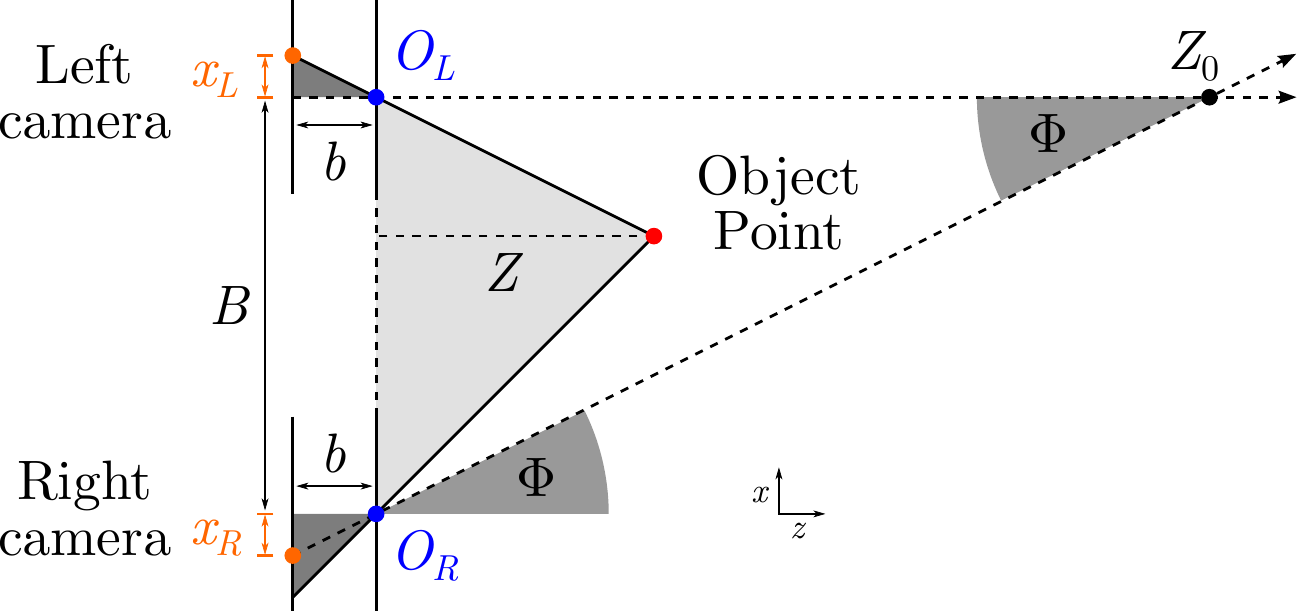}
	\caption[Stereo triangulation with non-parallel cameras]{Stereo triangulation scheme with non-parallel cameras where sensors are seen to be coplanar. $\Phi$ denotes the tilt angle of the right camera's main lens $O_R$ as related to that of the left camera $O_L$.}
	\label{fig:stereoscopyAngle}
\end{figure}
Taking the left camera as the orientation reference, the right lens $O_R$ is seen to be tilted as shown in Fig.~\ref{fig:stereoscopyAngle}. %
In this case, perspective image rectification is commonly employed to correct for non-coplanar stereo vision setups~\citep{Burger:2009:PDI:1529928}. 
\citeauthor*{IOCCHI}~(\citeyear{IOCCHI}) concludes that optical axes intersect in a point $Z_0$ as both axes lie on the $x,z$ plane if angle rotation occurs around the $y$-axis, whereas image planes of both cameras are still seen to be parallel. In traditional stereo vision, this yields deviations such that \citeauthor*{IOCCHI}'s~(\citeyear{IOCCHI}) method serves as a first-order approximation for small angle rotations in the absence of image processing. As demonstrated in Section~\ref{ssec:virtualCamera}, this approach, however, is suitable for our plenoptic triangulation model where imaginary sensor planes of virtual cameras are coplanar, whilst their optical axes may be non-parallel. Let $\Phi$ be the rotation angle, then laws of trigonometry allow to put
\begin{align}
	Z_0=\frac{B}{\tan(\Phi)} \label{eq:Z0}
\end{align}
and 
\begin{align}
	Z = \cfrac{b \times B}{\Delta x + \cfrac{b \times B}{Z_0}} \label{eq:Ztilt_Z0}
\end{align}
which may be shortened to
\begin{align}
	Z = \cfrac{b \times B}{\Delta x + b \times \tan(\Phi)} \label{eq:Ztilt} \,
\end{align}
after substituting for $Z_0$. This approximation suffices to estimate the depth $Z$ for small rotation angles $\Phi$ in stereoscopic systems without the need of an image rectification. %
\pagebreak
\section{SPC Ray Model}
\label{sec:2}
\begin{figure}[H]
	\centering
	\begin{minipage}[t]{\linewidth}
		\centering
		\includegraphics[width=\figWidthOneCols\linewidth]{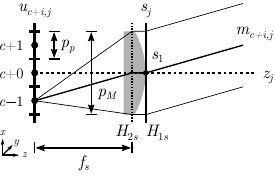}
		\subcaption{\label{fig:lensComponents:a}}
	\end{minipage}
	\\
	\begin{minipage}[t]{\linewidth}
		\centering
		\includegraphics[width=\linewidth]{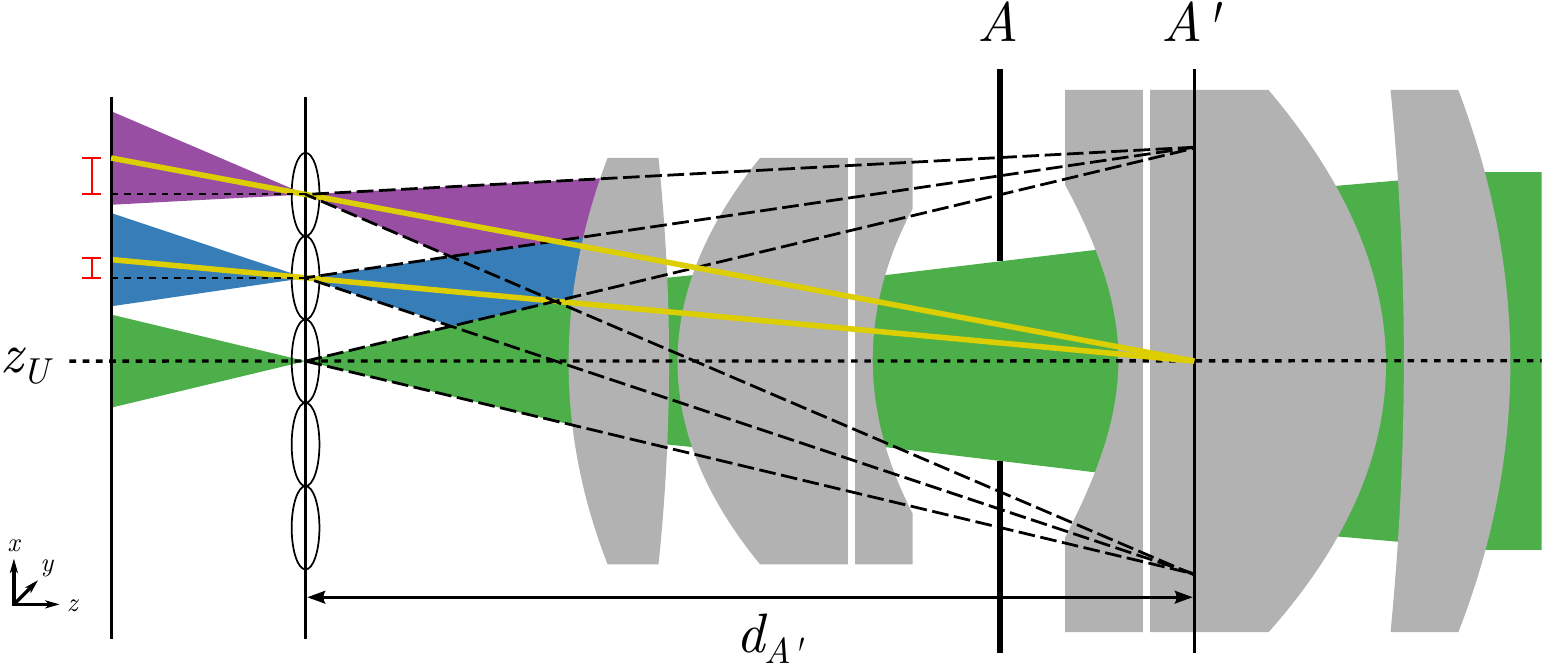}
		\subcaption{\label{fig:lensComponents:b}}
	\end{minipage}
	\caption{Lens components of plenoptic camera~\citep{Hahne:16:OPEX} depicting a micro lens $s_j$ with pitch size $p_M$ in \subref{fig:lensComponents:a} and an objective lens with exit pupil $A'$ in \subref{fig:lensComponents:b}. A chief ray $m_{c+i,j}$ pierces through the micro lens centre and sensor sampling positions $c+i$ which are separated by pixel width $p_p$.  Chief rays originate from the exit pupil centre $A'$ and arrive at \textit{Micro Image Centres}~(MICs) where red coloured crossbars signify gaps between MICs and respective micro lens optical axes. It can be seen that red crossbars grow towards image edges. \label{fig:lensComponents}}
\end{figure}

To conceptualise a light field ray model for an SPC, we start tracing rays from the sensor side to the object space. For simplification, we consider chief rays only and follow their path from each sensor's pixel centre at micro image domain $u$ to the optical centre of its corresponding micro lens $s_j$ with lens index $j$. 
Figure~\ref{fig:lensComponents} visualises chief rays travelling through a micro lens and the objective lens indicating \textit{Micro Image Centres}~(MICs). With the aid of ray geometry, an MIC is found by a chief ray connecting an optical centre of a micro lens with that of the main lens. %
MICs play a key role in realigning a light field from an SPC and are locally obtained by $c~=~\sfrac{(M - 1)}{2}$, where $M$ indicates one-dimensional {(1-D)} micro image resolutions, which are seen to be consistent. Discrete micro image points in the horizontal direction are then indexed by ${c+i}$, where ${i \in [-c,c]}$ such that \mbox{1-D} micro image samples are given as $u_{c+i,j}$. 
\par
In earlier publications~\citep{Hahne:14:IEEE,Hahne:14:OPEX}, it was assumed that MICs lie on the optical axes of corresponding micro lenses. However, it has been argued that this assumption would only be true if the distance between objective lens and MLA would be infinitely large~\citep{DANSPHD}. Due to the finite separation, MICs are displaced from their micro lens optical axes. A more accurate approach in estimating MIC positions is to model chief rays in a way that they connect optical centres of micro and main lenses~\citep{DANSCAL}. %
In Fig.~\ref{fig:lensComponents:b} we further refine this hypothesis by regarding the centre of an exit pupil $A'$ to be the origin from which MIC chief rays arise. %
Detecting MICs correctly is essential for our geometrical light ray model because MICs serve as reference points in the viewpoint image synthesis. 
%
\par
%
\begin{figure}[ht]
	\centering
	\includegraphics[width=\figWidthOneCols\linewidth]{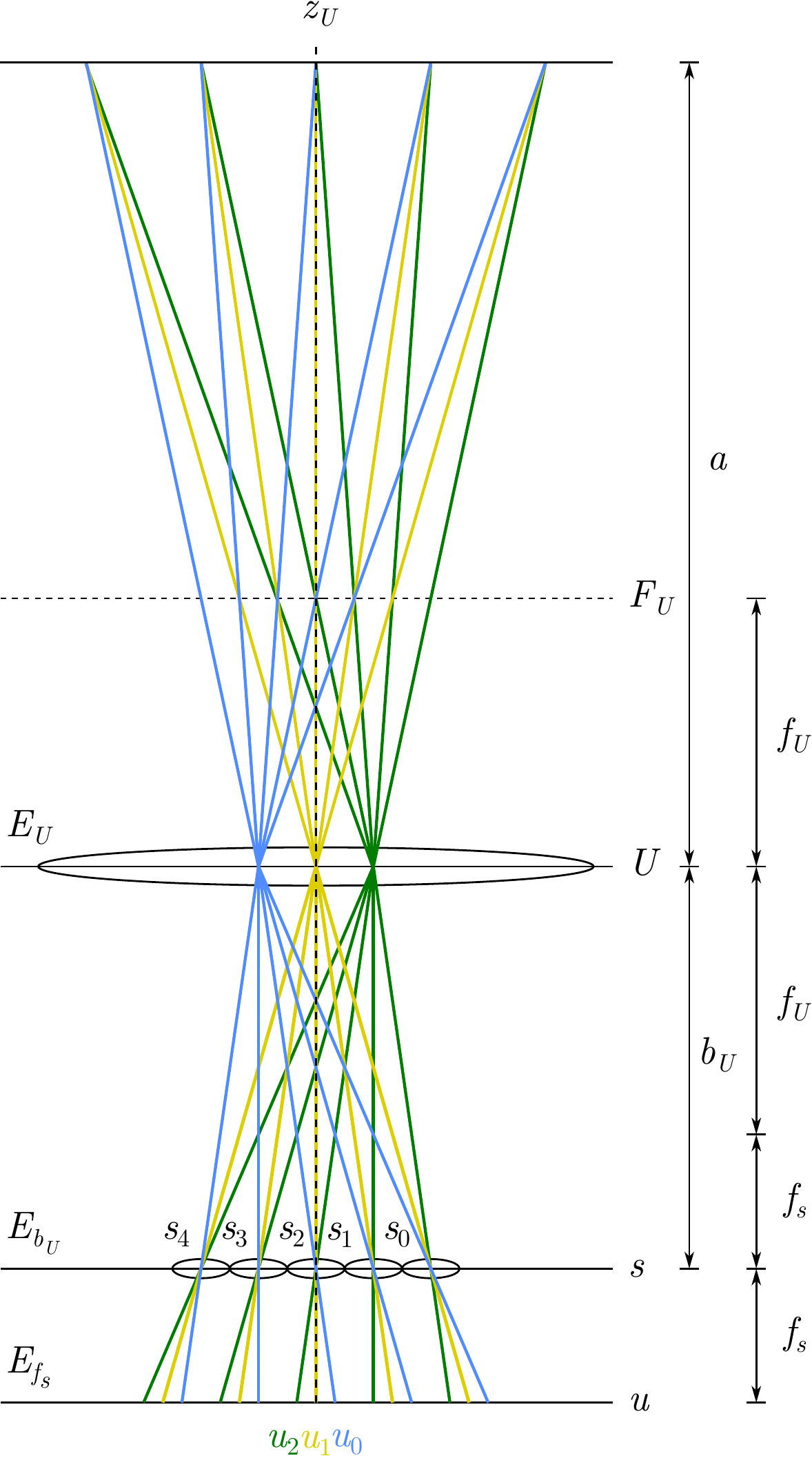}
	\caption[SPC ray model]{Illustration of the SPC ray model~\citep{Hahne:16:OPEX}, where MICs can be found by connecting the optical centre of the main lens with that of each micro lens and extending these rays (highlighted in yellow) until they reach the sensor. Here, the main lens is modelled as a thin lens such that entrance and exit pupil are in line with principal planes. %
	}
	\label{fig:refinedModel}
\end{figure}
%
%
Figure~\ref{fig:refinedModel} depicts our more advanced model that combines statements made about light rays' paths in an SPC. For clarity, the main lens $U$ is depicted as a thin lens meaning that the exit pupil centre coincides with the optical centre. However, the distinction is maintained in the following. %
%
\subsection{Viewpoint Extraction}
\begin{figure}[ht]
	\centering
	\begin{minipage}[t]{\linewidth}
		\centering
		\includegraphics[width=\figWidthOneCols\linewidth]{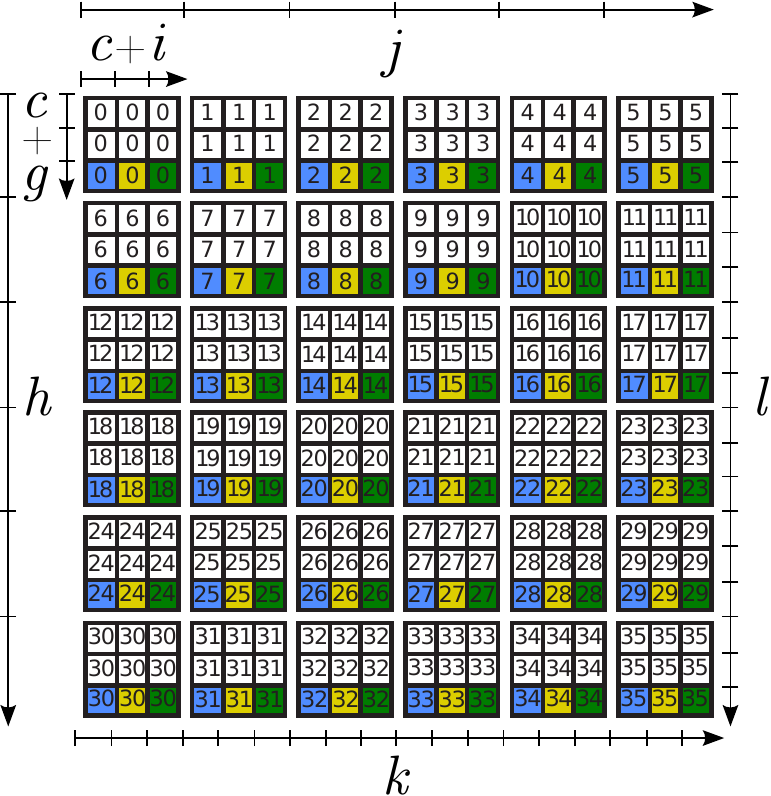}
		\subcaption{\label{fig:vpExtraction:a}}
	\end{minipage}
	\\
	\vspace{.25cm}
	\begin{minipage}[t]{\linewidth}
		\centering
		\includegraphics[width=\figWidthOneCols\linewidth]{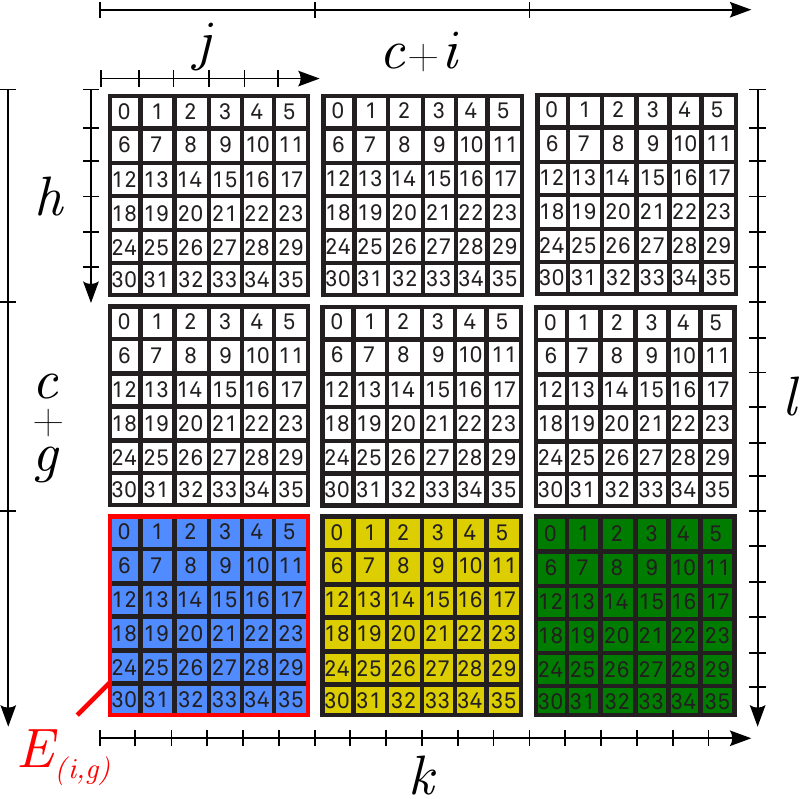}
		\subcaption{\label{fig:vpExtraction:b}}
	\end{minipage}
	\caption[Multiple sub-aperture image extraction]{Multiple sub-aperture image extraction with a calibrated raw image in~\subref{fig:vpExtraction:a} as obtained by an SPC and extracted \mbox{2-D} sub-aperture images $E_{(i,g)}$ in~\subref{fig:vpExtraction:b} where each colour represents a different perspective view. Note that the above figures consider a 180$\degree$ image rotation by the sensor to compensate for main lens image rotation. Micro image samples are indexed by $\left[s_j \, , \, t_h\right]$ and pixels within micro images by $\left[u_{c+i} \, , \, v_{c+g}\right]$ with $M=3$. Coordinates $\left[u_{c+i} \, , \, v_{c+g}\right]$ index viewpoint images and $\left[s_j \, , \, t_h\right]$ their related spatial pixels.}
	\label{fig:vpExtraction}
\end{figure}

It has been shown in~\citeauthor{AW}~\citeyearpar{AW}, \citeauthor{NG}~\citeyearpar{NG}, \citeauthor{DANSPHD}~\citeyearpar{DANSPHD}, \citeauthor{bok:eccv14}~\citeyearpar{bok:eccv14} that extracting viewpoints from an SPC can be attained by collecting all pixels sharing the same respective micro image position. To comply with provided notations, a \mbox{1-D} sub-aperture image $E_{i}\left[s_j\right]$ with viewpoint index $i$ is computed with
\begin{align}
	E_{i}\left[s_j\right] = E_{f_s}\left[s_j \, , \, u_{c+i}\right] \label{eq:vpExtract}
\end{align}
where $u$ and $c$ have been omitted in the subscript of $E_{i}$ since $i$ is a sufficient index for sub-aperture images in the \mbox{1-D} row. Equation~(\ref{eq:vpExtract}) implies that the effective viewpoint resolution equals the number of micro lenses. Figure~\ref{fig:vpExtraction} depicts the reordering process producing \mbox{2-D} sub-aperture images $E_{(i,g)}$ by means of index variables $\left[s_j \, , \, t_h\right]$ and $\left[u_{c+i} \, , \, v_{c+g}\right]$ for spatial and directional domains, respectively. As can be seen from colour-highlighted pixels, samples at a specific micro image position correspond to the respective viewpoint location in a camera array. %
\par
Since raw SPC captures do not naturally feature the $E_{f_s}\left[s_j \, , \, u_{c+i}\right]$ index notation, it is convenient to define an index translation formula considering the light field photograph to be of two regular sensor dimensions, $\left[x_k \, , \, y_l\right]$ as taken with a conventional sensor. In the horizontal dimension indices are converted by
\begin{align}
	k &=j \times M+c+i \, , \label{eq:translate}
\end{align}
which means that $\left[x_k\right]$ is formed by
\begin{align}
	\left[x_k\right] &= \left[x_{j \times M+c+i}\right] = \left[s_j \, , \, u_{c+i}\right] \label{eq:translate2} \, .
\end{align}
bearing in mind that $M$ represents the 1-D micro image resolution. Similarly, the vertical index translation may be
\begin{align}
	l &= h \times M+c+g \label{eq:translate3}
\end{align}
and therefore
\begin{align}
	\left[y_l\right] &= \left[y_{h \times M+c+g}\right] = \left[t_h \, , \, v_{c+g}\right] \label{eq:translate4} \, .
\end{align}
These definitions comply with Fig.~\ref{fig:vpExtraction} and enable to apply our {4-D} light field notation $\left[s_j \, , \, u_{c+i} \, , \, t_h \, , \, v_{c+g}\right]$ to conventionally \mbox{2-D} sampled representations $\left[x_k \, , \, y_l\right]$ with $k~\in~\left[0, \, K\right)$ and $l~\in~\left[0, \, L\right)$. 
To apply the proposed ray model and image process, the captured light field has to be calibrated and rectified such that the centroid of each micro image coincides with the centre of a central pixel. This requires an image interpolation with sub-pixel precision\if in MIC detection and viewpoint extraction\fi, which was first pointed out by \citeauthor{Cho:2013}~\citeyearpar{Cho:2013} and confirmed by \citeauthor{DANSCAL}~\citeyearpar{DANSCAL}.
\par %
%
\subsection{Virtual Camera Array}
\label{ssec:virtualCamera}
In the previous section, it was shown how to render {\it multi-views} from SPC photographs by means of the proposed ray model. Because a \mbox{4-D} plenoptic camera image can be reorganised to a set of {\it multi-view} images as if taken with an array of cameras, it is supposed that each of these images has an optical centre of a so-called virtual camera with a distinct location. The localisation of such is, however, not obvious. This problem was first recognised and addressed in publications by our research group~\citep{Hahne:14:IEEE,Hahne:14:OPEX} but, however, lacked of experimental verification. As a starting point, we deploy ray functions that proved to be viable to pinpoint refocused SPC image planes~\citep{Hahne:16:OPEX} and further refine the model by finding intersections along the entrance pupil. Once theoretical positions of virtual cameras are derived, we examine in which way the well established concept of stereo triangulation (see Section~\ref{sec:1}) applies to the proposed SPC ray model. \par%
In order to geometrically describe rays in the light field, we first define the height of optical centres $s_j$ in the MLA by
\begin{align}
s_j &= (j-o) \times p_{M} \label{eq:rayTracingEq1} 
\end{align}
with $o=\sfrac{(J-1)}{2}$ as the index of the central micro lens where $J$ is the overall number of micro lenses in the horizontal direction. Geometrical MIC positions are denoted as $u_{c,j}$ and can be found by tracing main lens chief rays travelling through the optical centre of each micro lens. This is calculated by
\begin{align}
u_{c,j} &= \frac{s_j}{d_{A'}} \times f_s + s_j \, , \label{eq:rayTracingEq5}
\end{align}
where $f_s$ is the micro lens focal length and $d_{A'}$ is the distance from MLA to exit pupil of the main lens, which is illustrated in Fig.~\ref{fig:lensComponents:b}. Micro image sampling positions that lie next to MICs can be acquired by a corresponding multiple $i$ of the pixel pitch~$p_p$ as given by %
\begin{align}
u_{c+i,j} &= u_{c,j} + i \times p_p \, . \label{eq:rayTracingEq6}
\end{align}
Chief ray slopes $m_{c+i,j}$ that impinge at micro image positions $u_{c+i,j}$ can be acquired by
\begin{align}
m_{c+i,j} = \frac{s_j - u_{c+i,j}}{f_s} \, . \label{eq:rayTracingEq7}
\end{align}
Let $b_U$ be the objective's image distance, then a chief ray's intersection at the refractive main lens plane $U_{i,j}$ is given by
\begin{align}
U_{i,j} &= m_{c+i,j} \times b_U + s_j \, . \label{eq:rayTracingEq8}
\end{align}
where $c$ has been left out in the subscript of $U_{i,j}$ as it is a constant and will be omitted in following ray functions for simplicity. The spacing between principal planes of an objective lens will be taken into account at a later stage. %
\par
The main lens works as a refracting element such that chief rays possess different slopes in object space. A ray slope in object space can be modelled using auxiliary paths to points $F_{i,j}$ located along the main lens focal plane $F$, where corresponding chief rays pass through. From Gaussian optics it follows that the position $F_{i,j}$ is given by
\begin{align}
F_{i,j} &= m_{c+i,j} \times f_U \, , \label{eq:rayTracingEq9}
\end{align}
governing the slope $q_{i,j}$ of a chief ray in object space, which is obtained by
\begin{align}
q_{i,j} &= \frac{F_{i,j} - U_{i,j}}{f_U} \label{eq:rayTracingEq10}
\end{align}
as it depends on the intersections at refractive main lens plane $U$, focal plane $F_U$ and the chief ray's travelling distance $f_U$. %
With reference to preliminary remarks, an object ray path may be provided as a linear function $\widehat{f}_{i,j}$ of the depth $z$, which is written as
\begin{align}
\widehat{f}_{i,j}(z) &= q_{i,j} \times z+U_{i,j} \, \, ,\quad z \in \left[U,\infty\right) \, . \label{eq:rayTracingEq11}
\end{align}
\par
As the name suggests, sub-aperture images are created at the main lens aperture. To investigate ray positions at the aperture, we introduce the exit and entrance pupil as respective geometrical equivalents to the proposed model, which have not been considered in \citep{Hahne:14:IEEE}. An obvious attempt would be to locate a baseline $B_{A'}$ at the exit pupil, which is found by
\begin{align}
	B_{A'} = m_{c+i,j} \times d_{A'} \, , \label{eq:exitPupilBaseline}
\end{align}
where $m_{c+i,j}$ is obtained from Eq.~(\ref{eq:rayTracingEq7}). 
Practical applications of an image-side baseline $B_{A'}$ are unclear at this stage. \par
\begin{figure}[H]
	\centering
	\includegraphics[width=\figWidthOneCols\linewidth]{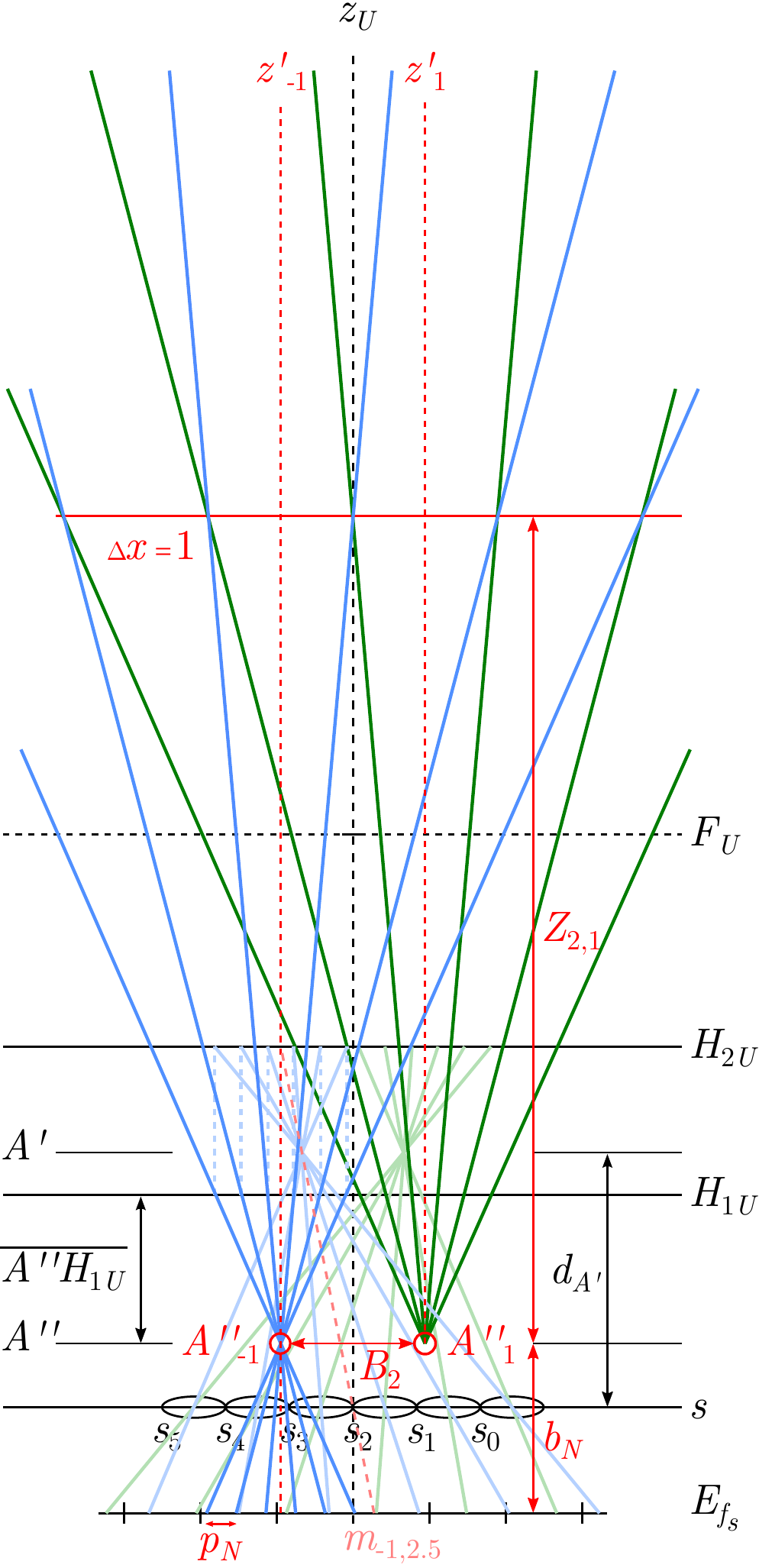}
	\caption[SPC model triangulation with $b_U=f_U$]{SPC model triangulation with $b_U=f_U$ and principal planes $H_{1U}$, $H_{2U}$ just as the exit $A'$ and entrance pupil plane $A''$. Red circles next to $A''_{i}$ indicate virtual camera positions. Note that virtual cameras $A''_{-1}$ and $A''_1$ are separated by gap $G=2$ yielding baseline $B_2$.}
	\label{fig:baseline:1}
\end{figure}

However, the baseline at the entrance pupil $A''$ is a much more valuable parameter when determining an object distance via triangulation in an SPC. Figure~\ref{fig:baseline:1} offers a closer look at our light field ray model by also showing principal planes $H_{1U}$ and $H_{2U}$. %
There, it can be seen that all rays having $i$ in common (e.g. blue rays) geometrically converge to the entrance pupil $A''$ and diverge from the exit pupil $A'$. Intersecting chief rays at the entrance pupil can be seen as indicating object-side-related positions of virtual cameras $A''_{i}$. \par
\begin{figure}[H]
	\centering
	\includegraphics[width=\figWidthOneCols\linewidth]{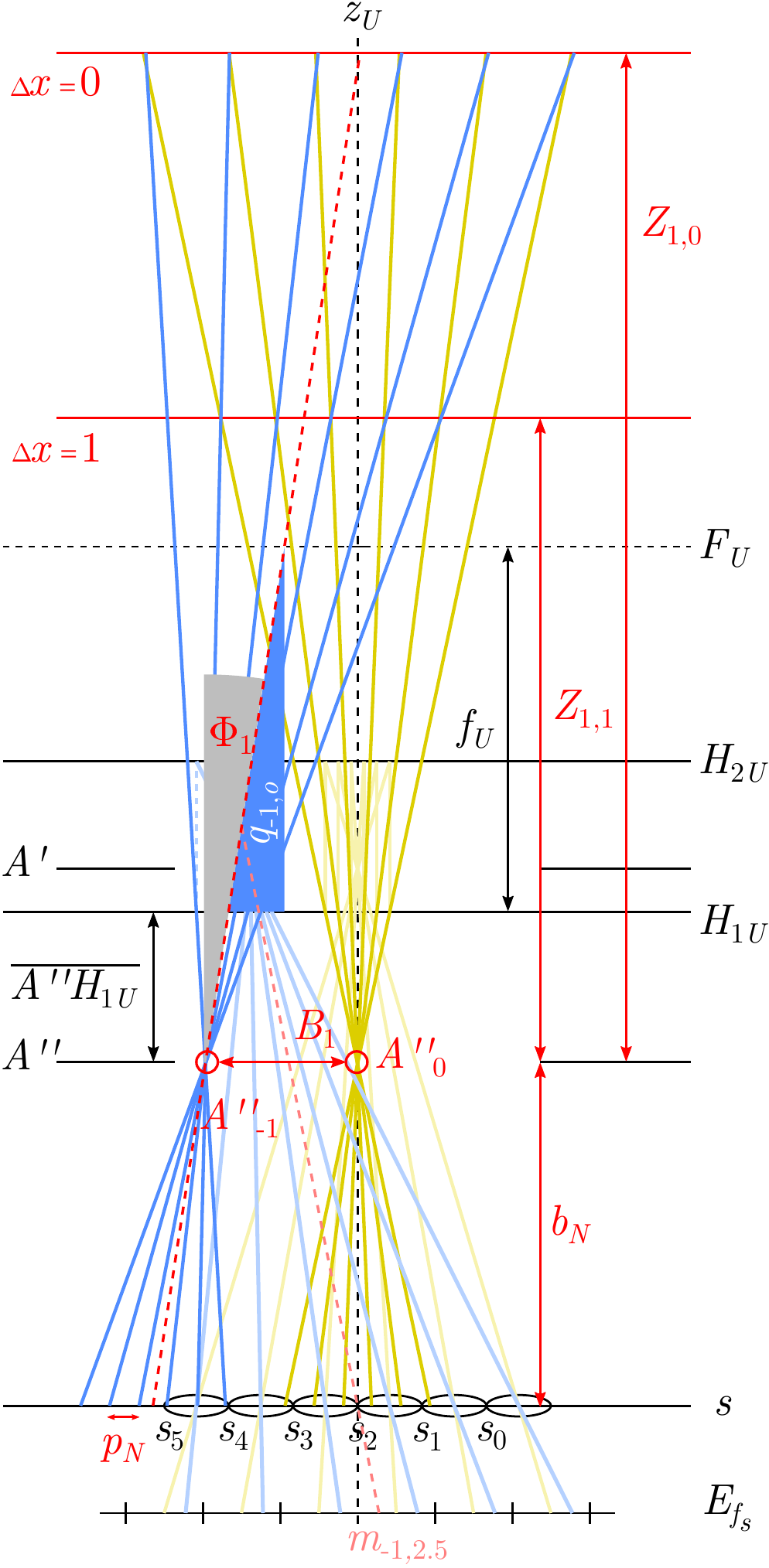}
	\caption[SPC model triangulation with $b_U>f_U$]{SPC model triangulation with $b_U>f_U$. Red circles next to $A''_{i}$ indicate virtual camera positions. Note that the gap $G=1$ and therefore $B_1$ and $\Phi_1$.}
	\label{fig:baseline:2}
\end{figure}

The calculation of virtual camera positions $A''_{i}$ is provided in the following.
By taking object space ray functions $\widehat{f}_{i,j}(z)$ from Eq.~(\ref{eq:rayTracingEq11}) for two rays with different $j$ but same $i$ and setting them equal as given by
\begin{align}
q_{i,o}\times z + U_{i,o} = q_{i,o+1}\times z + U_{i,o+1} \, , \quad z \in \left(-\infty \, , \, \infty \, \right) \, , 
\label{eq:BaselineSlope}
\end{align}
we can solve for the equation system which yields a distance $\overline{A''H_{1U}}$ from entrance pupil $A''$ to object-side principal plane $H_{1U}$ (see Fig.~\ref{fig:baseline:1}). Recall that the index for the central micro lens $s_j$ is found by $j=o=\sfrac{(J-1)}{2}$ where $o$ defines the image centre offset. %
%
The object-side-related position of a virtual camera $A''_i$ can be acquired by
\begin{align}
	A''_i = q_{i,o} \times \overline{A''H_{1U}} + U_{i,o} \, . \label{eq:virtualCamera}
\end{align}
With this, a baseline $B_{G}$ that spans from one $A''_{i}$ to another by gap $G$ can be obtained as follows
\begin{align}
	B_{G} = A''_{i} + A''_{i+G} \, . \label{eq:baseline}
\end{align}
%
For example, a baseline $B_1$ ranging from $A''_{0}$ to $A''_{1}$ is identical to that from $A''_{-1}$ to $A''_{0}$. This relies on the principle that virtual cameras are separated by a consistent width. To apply the triangulation concept, rays are virtually extended towards the image space by
\begin{align}
	N_{i,j} = -q_{i,j} \times b_N + A''_{i} \, ,
\end{align}
where $b_N$ is an arbitrary scalar which can be thought of as a virtual image distance and $N_{i,j}$ as a spatial position at the virtual image plane of a corresponding sub-aperture. The scalable variable $b_N$ linearly affects a virtual pixel pitch $p_N$, which is found by
\begin{align}
	p_N = \big|N_{i,o} - N_{i,o+1}\big| \, .
\end{align}
Setting $b_U=f_U$ aligns optical axes $z'_i$ of virtual cameras to be parallel to the main optical axis $z_U$ (see Fig.~\ref{fig:baseline:1}). 
For all other cases where $b_U\neq f_U$ (e.g. Fig.~\ref{fig:baseline:2}), the rotation angle $\Phi_i$ of a virtual optical axis $z'_i$ is obtained by
\begin{align}
	\Phi_i = \arctan{\left(q_{i,o}\right)} \, . \label{eq:phi}
\end{align}
The relative tilt angle $\Phi_G$ from one camera to another can be calculated with
\begin{align}
	\Phi_{G} = \Phi_{i} + \Phi_{i+G} \, , \label{eq:phi2}
\end{align}
which completes the characterisation of virtual cameras. \par
Figure~\ref{fig:baseline:2} visualises chief rays' paths in the light field when focusing the objective lens such that $b_U>f_U$. In this case, $z'_i$ intersects with $z_U$ at the plane at which the objective lens is focusing. Objects placed at this plane possess a disparity $\Delta x = 0$ and thus are expected to be located at the same relative \mbox{2-D} position in each sub-aperture image. As a consequence, objects placed behind the $\Delta x = 0$ plane expose negative disparity. \par
%
Establishing the triangulation in an SPC allows object distances to be retrieved just as in a stereoscopic camera system. On the basis of Eq.~(\ref{eq:Ztilt}), a depth distance $Z_{G,\Delta x}$ of an object with certain disparity $\Delta x$ is obtained by
\begin{align}
Z_{G,\Delta x} = \frac{b_N \times B_{G}}{\Delta x \times p_N + b_N \times \tan \left(\Phi_G\right)} \label{eq:SPCtriangulation}
\end{align}
and can be shortened to
\begin{align}
Z_{G,\Delta x} = \frac{b_N \times B_{G}}{\Delta x \times p_N} \, , \quad \text{if} \, \, \Phi_G = 0 \label{eq:shortSPCtriangulation}
\end{align}
which is only the case where $b_U=f_U$. One may notice that Eq.~\eqref{eq:shortSPCtriangulation} is an adapted version of the well-known triangulation equation given in Eq.~\eqref{eq:classicTria}. \par
%
%
\section{Validation}
\label{sec:3}
\begin{figure}[h]
	\centering
		\begin{minipage}[t]{.495\linewidth}
			\centering
			\includegraphics[width=.95\linewidth,trim= 0cm 0cm 0cm 6cm, clip=true]{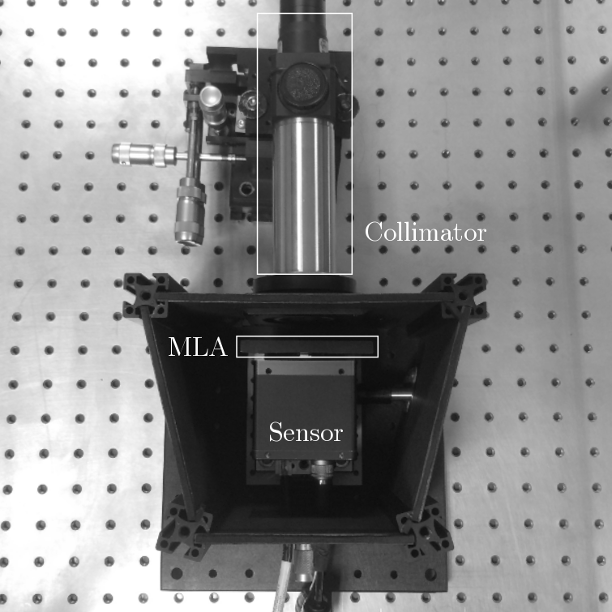}
		\end{minipage}
		\begin{minipage}[t]{.495\linewidth}
			\centering
			\includegraphics[width=.95\linewidth,trim= 0cm 4cm 0cm 2cm, clip=true]{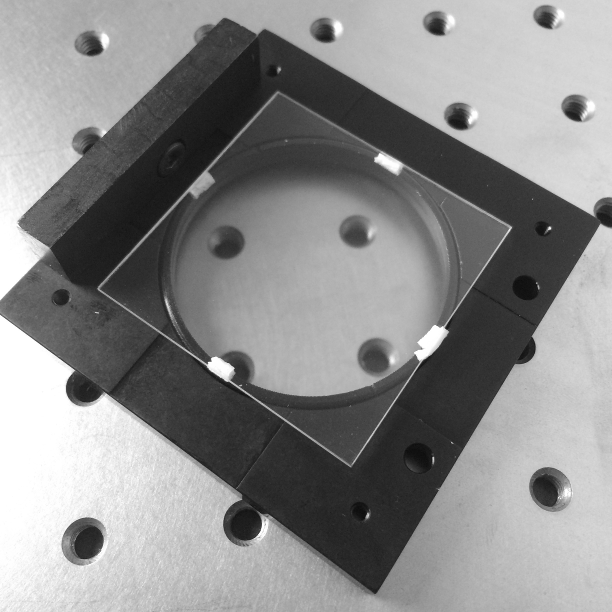}
		\end{minipage}	
	\caption[Photographs from MLA setup]{Photographs from our custom-built camera with camera body and collimator (left) and MLA fixation (right). \label{fig:mlaSetup}}
\end{figure}

We deploy a custom-made plenoptic camera containing a \textit{full frame} sensor with 4008 x 2672 active image resolution and $p_p~=~9~\mu$m pixel pitch. Photos of our camera are depicted in Fig.~\ref{fig:mlaSetup}. Details on the assembly and optical calibration of an SPC can be found in \citeauthor*{Hahne:Thesis}'s~thesis~(\citeyear{Hahne:Thesis}). Lens and MLA specifications are provided hereafter.%
\subsection{Lens Specifications}
Experimentations are conducted with two different micro lens designs, denoted as MLA~(\RM{1})~and~(\RM{2}), which can be found in Table~\ref{tab:MLA}. Input parameters relevant to the triangulation are $f_s$ and $p_m$. Besides this, Table~\ref{tab:MLA} provides the lens thickness $t_s$, refractive index $n$, radii of curvature $R_{s1}$, $R_{s2}$ and principal plane distance $\overline{H_{1s}H_{2s}}$. 
The number of micro lenses in our MLA amounts to 281 x 188 for horizontal and vertical dimensions, respectively. These values allow for modelling the micro lenses in an optical design software. \par%
\begin{table}[ht]
	\centering
	\caption[Micro lens specifications for $\lambda=550$~nm]{Micro lens specifications for $\lambda=550$~nm.}
	\renewcommand{\arraystretch}{1}
	\resizebox{\tabWidthTwoCols\columnwidth}{!}{%
	\begin{tabular}{c c c c c c c c c c }
		\toprule
		MLA & $f_s$ & $p_M$ & $t_s$ & $n(\lambda)$ & $R_{s1}$ & $R_{s2}$ & $\overline{H_{1s}H_{2s}}$ \\ 
		\midrule
		(\RM{1}) & 1.25~mm & $125~\mu $m & 1.1~mm & 1.5626 & 0.70325 & -$\infty$ & 0.396
		~mm \\
		(\RM{2}) & 2.75~mm & $125~\mu $m & 1.1~mm & 1.5626 & 1.54715 & -$\infty$ & 0.396
		 ~mm \\ 
		\bottomrule
	\end{tabular}
	\label{tab:MLA}
	}
\end{table}\noindent
\par
It is well known that the focus ring of today's objective lenses moves a few lens groups whilst others remain static which, in consequence, changes the lens system's cardinal points. To prevent this and simplify the experimental setup, we only shift the plenoptic sensor away from the main lens to vary its image distance $b_U$ by keeping the focus ring at infinity. In doing so, we assure cardinal points remain at the same relative position. However, the available space in our customised camera constrains the sensor's shift range to an overall focus distance of $d_f \approx$~4~m where $d_f$ is the distance from the MLA's front vertex to the plane that the main lens is focused on. For this reason, we examine two focus settings $(d_f \to \infty~\text{and}~d_f \approx~4~\text{m})$ in the experiment. %
To acquire the main lens image distance $b_U$, we employ the thin lens equation and solve for $b_U$ as given by
\begin{align}
	b_U &= \left(\frac{1}{f_U} - \frac{1}{a_U}\right)^{-1} \, ,
\end{align}
with $a_U=d_f - b_U - \overline{H_{1U}H_{2U}}$ as the object distance. 
After substituting for $a_U$, however, it can be seen that $b_U$ is an input and output parameter at the same time, which turns out to be a typical \textit{chicken-and-egg} case. To treat this problem, we define the initial image distance to be the focal length ($b_U:=f_U$) and substitute the resulting $b_U$ for the input variable afterwards. This procedure is iterated until both values are the same. %
Objective lenses are denoted as $f_{193}$, $f_{90}$ and $f_{197}$ with index numbers representing focal lengths in millimetres. The lens designs for $f_{193}$ and $f_{90}$ were found in~\citep{CALDWELL:2000,yanagisawa1990optical} whilst $f_{197}$ is obtained experimentally using the technique provided by \cite{TRIOPTICS}. Table~\ref{tab:bU} lists calculated image, exit pupil and principal plane distances for the main lenses. It is noteworthy that all parameters are provided with respect to 550~nm wavelength. Precise focal lengths $f_U$ are found in the image distance column at the infinity focus row. %
\begin{table}[ht]
	\small
	\centering
	\caption[Main lens parameters]{Main lens parameters.}
	\label{tab:bU}
	\renewcommand{\arraystretch}{1}
	\resizebox{\linewidth}{!}{%
	\begin{tabular}{ 
	    |c
	    |S[table-format=3.4]
	    |S[table-format=2.4]
	    |S[table-format=3.4]
	    |S[table-format=3.4]
	    |S[table-format=2.4]
	    |S[table-format=3.4]|
	}
		\hline
		\textbf{Focus} & \multicolumn{3}{c|}{\textbf{Image distance}} & \multicolumn{3}{c|}{\textbf{Exit pupil position}} \\
		\hline
		\multirow{2}{*}{$d_f$} & \multicolumn{3}{c|}{$b_U$ [mm]} & \multicolumn{3}{c|}{$d_{A'}$ [mm]} \\
		\cline{2-7}
		& $f_{193}$ & $f_{90}$ & $f_{197}$ & $f_{193}$ & $f_{90}$ & $f_{197}$ \\ 
		\hline
		$\infty$ & 193.2935 & 90.4036 & 197.1264 & 111.0324 & 85.1198 & 100.5000 \\ 
		4~m & {--} & {--} & 208.3930 & {--} & {--} & 111.7666 \\ 
		3~m & 207.3134 & 93.3043 & {--} & 125.0523 & 88.0205 & {--} \\ 
		1.5~m & 225.8852 & 96.6224 & {--} & 143.6241 & 91.3386 & {--} \\ 
		\hline
	\end{tabular}
	}
	\newline
	\vspace*{.5 cm}
	\newline
	\resizebox{.215\textwidth}{!}{%
		\begin{tabular}{ %
			|S[table-format=3.4]
			|S[table-format=2.4]
			|S[table-format=3.4]|
			}
			\hline
			\multicolumn{3}{|c|}{\textbf{Principal plane separation}} \\
			\hline
			\multicolumn{3}{|c|}{$\overline{H_{1U}H_{2U}}$ [mm]} \\
			\hline
			$f_{193}$ & $f_{90}$ & $f_{197}$ \\
			\hline
			{-65.5563} & {-1.2273} & {147.4618} \\
			\hline
		\end{tabular}
	}
\end{table}
%
\subsection{Experiments}
%
%
To verify claims made about SPC triangulation, experiments are conducted as follows. Baselines and tilt angles are estimated based on \Cref{eq:baseline,eq:phi2} using parameters given in \Cref{tab:MLA,tab:bU}. Thereof, we compute object distances from Eq.~\eqref{eq:SPCtriangulation} for each disparity and place real objects at the calculated distances. 
Experimental validation is achieved by comparing predicted baselines with those obtained from disparity measurements. %
The extraction of a disparity map from an SPC requires at least two sub-aperture images that are obtained using Eq.~(\ref{eq:vpExtract}). Disparity maps are calculated by block matching with the \textit{Sum of Absolute Differences}~(SAD) method using an available implementation~\citep{Abbeloos:Thesis:2010,Abbeloos:Software:2012}. %
To measure baselines, Eq.~(\ref{eq:SPCtriangulation}) has to be rearranged such that
\begin{align}
B_{G} = 
\frac{Z_{G,\Delta x} \times \left(\Delta x \times p_N + b_N \times \tan \left(\Phi_G\right)\right)}{b_N} \, . \label{eq:SPCtriangulationBaseline}
\end{align}
This formula can also be written as
\begin{align}
\Phi_G = \arctan\left(\frac{\frac{B_{G} \times b_N}{Z_{G,\Delta x}} - \Delta x \times p_N}{b_N}\right) \, , \label{eq:SPCtriangulationTiltAngle}
\end{align}
which yields a relative tilt angle $\Phi_G$ in radians that can be converted to degrees by multiplication by $180/\pi$. %
%
\par
Stereo triangulation experiments are conducted such that $B_4$ and $B_8$, just as $\Phi_4$ and $\Phi_8$, are predicted based on main lens $f_{197}$ and MLA (\RM{2}) with $d_f \to \infty$ and $d_f \approx 4$~m focus setting. Real objects were placed at selected depth distances $Z_{G,\Delta x}$ calculated from this setup. %
\begin{figure}[H]
	\centering
	\begin{minipage}[t]{\columnwidth}
		\centering
		\raisebox{-.5\height}{\includegraphics[width=2.215278in]{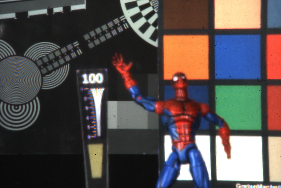}}
		\\[.17cm]
		\subcaption{Reference image $E_{(0,0)}$ where $d_f \to \infty$\label{fig:Base:Inf}}
		\vspace{.4cm}
	\end{minipage}
	\begin{minipage}[t]{\columnwidth}
		\centering
		\raisebox{-.5\height}{
		\begin{tikzpicture}	
			\begin{axis}[%
			width=2.204861in,
			height=1.475138in,
			at={(0.853125in,0.824931in)},
			scale only axis,
			axis on top,
			separate axis lines,
			every outer x axis line/.append style={black},
			every x tick label/.append style={font=\color{black}},
			xmin=0.5,
			xmax=281.5,
			tick align=outside,
			every outer y axis line/.append style={black},
			every y tick label/.append style={font=\color{black}},
			y dir=reverse,
			ymin=0.5,
			ymax=188.5,
			hide axis,
			colormap/jet,
			colorbar,
			colorbar style={
				ytick={0,1,2,3,4,5},
				separate axis lines,every outer x axis line/.append style={black},every x tick label/.append style={font=\color{black}},every outer y axis line/.append style={black},every y tick label/.append style={font=\color{black}}},
			point meta min=0,
			point meta max=5
			]
			\addplot [forget plot] graphics [xmin=0.5,xmax=281.5,ymin=0.5,ymax=188.5] {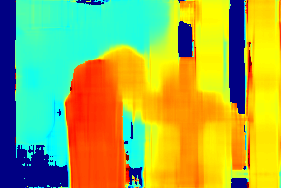};
			\end{axis}
		\end{tikzpicture}%
		}
		\subcaption{$\Delta x$ values from $E_{(-2,0)}$ and $E_{(2,0)}$ \label{fig:B4:Inf:m2p2}}
		\vspace{.4cm}
	\end{minipage}
	\begin{minipage}[t]{\columnwidth}
		\centering
		\begin{tikzpicture}		
			\begin{axis}[%
			width=2.204861in,
			height=1.475138in,
			at={(0.853125in,0.824931in)},
			scale only axis,
			axis on top,
			separate axis lines,
			every outer x axis line/.append style={black},
			every x tick label/.append style={font=\color{black}},
			xmin=0.5,
			xmax=281.5,
			tick align=outside,
			every outer y axis line/.append style={black},
			every y tick label/.append style={font=\color{black}},
			y dir=reverse,
			ymin=0.5,
			ymax=188.5,
			hide axis,
			colormap/jet,
			colorbar,
			colorbar style={
				separate axis lines,every outer x axis line/.append style={black},every x tick label/.append style={font=\color{black}},every outer y axis line/.append style={black},every y tick label/.append style={font=\color{black}}},
			point meta min=0,
			point meta max=9
			]
			\addplot [forget plot] graphics [xmin=0.5,xmax=281.5,ymin=0.5,ymax=188.5] {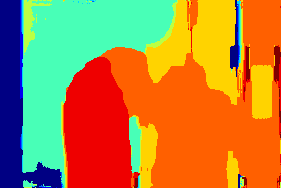};
			\end{axis}
		\end{tikzpicture}%
		\subcaption{$\Delta x$ values from $E_{(-4,0)}$ and $E_{(4,0)}$ \label{fig:B8:Inf:m4p4}}
		\vspace{.2cm}
	\end{minipage}
	\begin{minipage}[t]{\columnwidth}
		\centering
		\begin{tikzpicture}	
			\begin{axis}[%
			width=2.204861in,
			height=1.475138in,
			at={(0.853125in,0.824931in)},
			scale only axis,
			axis on top,
			separate axis lines,
			every outer x axis line/.append style={black},
			every x tick label/.append style={font=\color{black}},
			xmin=0.5,
			xmax=281.5,
			tick align=outside,
			every outer y axis line/.append style={black},
			every y tick label/.append style={font=\color{black}},
			y dir=reverse,
			ymin=0.5,
			ymax=188.5,
			hide axis,
			colormap/jet,
			colorbar,
			colorbar style={
				ytick={0,1,2,3,4,5},
				separate axis lines,every outer x axis line/.append style={black},every x tick label/.append style={font=\color{black}},every outer y axis line/.append style={black},every y tick label/.append style={font=\color{black}}},
			point meta min=0,
			point meta max=5
			]
			\addplot [forget plot] graphics [xmin=0.5,xmax=281.5,ymin=0.5,ymax=188.5] {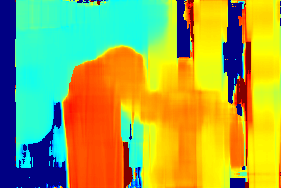};
			\end{axis}
		\end{tikzpicture}%
		\subcaption{$\Delta x$ values from $E_{(0,0)}$ and $E_{(4,0)}$ \label{fig:B4:Inf:0p4}}
		\vspace{.2cm}
	\end{minipage}
	\caption[Disparity maps from sub-aperture images $E_{(i,g)}$ with $b_U = f_U$]{Disparity maps from sub-aperture images $E_{(i,g)}$ with $b_U = f_U$. \subref{fig:Base:Inf} Central image $E_{(0,0)}$ containing $281$ by $188$ pixels; \subref{fig:B4:Inf:m2p2} Disp. map with $G=4$, $\text{max}\{\Delta x\} = 5$ and block size = 29; \subref{fig:B8:Inf:m4p4}  Disp. map with $G=8$, $\text{max}\{\Delta x\} = 9$ and block size = 39; \subref{fig:B4:Inf:0p4} Disp. map with $G=4$, $\text{max}\{\Delta x\} = 5$ and block size = 29.}
	\label{fig:f197Base:Inf}
\end{figure}
%
An exemplary sub-aperture image $E_{(i,g)}$ with infinity focus setting and related disparity maps is shown in Fig.~\ref{fig:f197Base:Inf}. A sub-pixel precise disparity measurement has been applied to \Cref{fig:B4:Inf:m2p2,fig:B4:Inf:0p4} as the action figure lies between integer disparities. %
It may be obvious that disparities in \Cref{fig:B4:Inf:m2p2,fig:B4:Inf:0p4} are nearly identical since both viewpoint pairs are separated by $G=4$, however placed at different horizontal positions. This justifies the claim that the spacing between adjacent virtual cameras is consistent. 
Besides, it is also apparent that objects at far distances expose lower disparity values and vice versa. Comparing \Cref{fig:B8:Inf:m4p4,fig:B4:Inf:m2p2} shows that a successive increase in the baseline $B_G$ implies a growth in the object's disparity values, an observation also found in traditional computer stereo vision. %
%
\begin{table}[H]
	\small
	\centering
	\captionsetup{justification=centering}
	\caption[Baseline results $B_G$ with infinity focus ($b_U=f_U$)]{Baseline results $B_G$ with infinity focus $(b_U=f_U)$.\label{tab:baseInfComb}}
	\begin{minipage}[b]{.48\linewidth}
		\centering
		\renewcommand{\arraystretch}{1.2}
		\subcaption{$B_4$ from \Cref{fig:B4:Inf:m2p2,fig:B4:Inf:0p4}\label{tab:baseInfComb:a}}
		\resizebox{\linewidth}{!}{%
		\begin{tabular}{ c c c }
			\toprule
			\textbf{$\Delta x$} & \textbf{$Z_{G,\Delta x}$~[cm]} & \multicolumn{1}{c}{\textbf{Measured $B_4$~[mm]}} \\
			\midrule
			2 & 203 & 2.5806 \\
			3 & 136 & 2.5806 \\
			3.5 & 116 & 2.5806 \\
			4 & 102 & 2.5806 \\
			\bottomrule
		\end{tabular}
		}
	\end{minipage}
	\hfill
	\begin{minipage}[b]{.48\linewidth}
		\centering
		\renewcommand{\arraystretch}{1.2}
		\subcaption{$B_8$ from Fig.~\ref{fig:B8:Inf:m4p4}\label{tab:baseInfComb:c}}
		\resizebox{\linewidth}{!}{%
		\begin{tabular}{ c c c }
			\toprule
			\textbf{$\Delta x$} & \textbf{$Z_{G,\Delta x}$~[cm]} & \multicolumn{1}{c}{\textbf{Measured $B_8$~[mm]}} \\
			\midrule
			4 & 203 & 5.1611 \\
			6 & 136 & 5.1611 \\
			7 & 116 & 5.1611 \\
			8 & 102 & 5.1611 \\
			\bottomrule
		\end{tabular}
		}
	\end{minipage}
	\\[4ex]
	\begin{minipage}[b]{1\linewidth}
		\centering
		\renewcommand{\arraystretch}{1.2}
		\subcaption{Comparison of predicted and measured $B_G$ where $d_f \to \infty$ \label{tab:baseInfComb:e}}
		\resizebox{\linewidth}{!}{%
		\begin{tabular}{ c c c c c }
			\toprule
			& & \textbf{Predicted} & \textbf{Avg. measured} & \textbf{Deviation} \\
			& \textbf{$B_G$} & $B_G$~[mm] & \multicolumn{1}{c}{$B_G$ [mm]} & \multicolumn{1}{c}{$ERR_{B_G}$~[\%]} \\
			\midrule
			\multirow{2}{*}{\textbf{Proposed}} & $B_4$ & 2.5806 & 2.5806 & 0.0000 \\
			& $B_8$ & 5.1611 & 5.1611 & 0.0000 \\
			\midrule
			\multirow{2}{*}{\cite{Hahne:14:IEEE,Hahne:14:OPEX}} & $B_4$ & 2.5806 & 12.0566 & -367.2090 \\ 
			& $B_8$ & 5.1611 & 24.1133 & -367.2090 \\ 
			\bottomrule
		\end{tabular}
		}
	\end{minipage}
	\captionsetup{justification=justified}
\end{table}
%
%
Table~\ref{tab:baseInfComb} lists baseline measurements and corresponding deviations with respect to the predicted baseline. This table is quite revealing in several ways. First, the most striking result is that there is no significant difference between baseline predictions and measurements using the model proposed in this paper. The reason for a 0~\% deviation is that objects are placed at the centre of predicted depth planes $Z_{G,\Delta x}$. An experiment conducted with random object positions would yield non-zero errors that do not reflect the model's accuracy, but rather our SPC's capability to resolve depth, which depends on MLA and sensor specification. Hence, such an experiment is only meaningful when evaluating the camera's depth resolution. A more revealing percentage error is obtained by a larger number of disparities, which in turn requires the baseline to be extended. These parameters have been maximised in our experimental setup making it difficult to further refine depth. 
To obtain quantitative error results, Subsection~\ref{ssec:sim} aims to benchmark proposed SPC triangulation with the aid of a simulation tool~\citep{Zemax}. \par%
A second observation is that our previous methods~\citep{Hahne:14:IEEE,Hahne:14:OPEX} yield identical baseline estimates, but fail experimental validation exhibiting significantly large errors in the triangulation. This is due to the fact that our previous model ignored pupil positions of the main lens such that virtual cameras were seen to be lined up on its front focal plane instead of its entrance pupil. Baseline estimates calculated according to a definition provided by \cite{jeon:cvpr15} further deviate from our results with $B_4 = 290.7293$~mm and $B_8 = 581.4586$~mm. As the authors disregard optical centre positions of the sub-aperture images, it is impossible to obtain distances via triangulation and assess results using percentage errors. \par
\begin{figure}[H]
	\centering
	\begin{minipage}[t]{\columnwidth}
		\centering
		\raisebox{-.5\height}{\includegraphics[width=2.215278in]{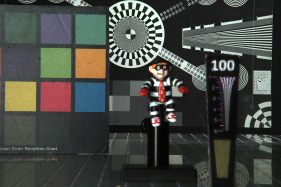}}
		\\[.17cm]
		\subcaption{Reference image $E_{(0,0)}$ where $d_f \approx 4$~m\label{fig:Base:4m}}
		\vspace{.4cm}
	\end{minipage}
	\begin{minipage}[t]{\columnwidth}
		\centering
		\raisebox{-.5\height}{
		\begin{tikzpicture}
			\begin{axis}[%
			width=2.204861in,
			height=1.467292in,
			at={(0.853125in,0.823646in)},
			scale only axis,
			axis on top,
			separate axis lines,
			every outer x axis line/.append style={black},
			every x tick label/.append style={font=\color{black}},
			xmin=0.5,
			xmax=281.5,
			tick align=outside,
			every outer y axis line/.append style={black},
			every y tick label/.append style={font=\color{black}},
			y dir=reverse,
			ymin=0.5,
			ymax=187.5,
			hide axis,
			colormap/jet,
			colorbar,
			colorbar style={
				ytick={0,1,2,3,4,5},
				separate axis lines,every outer x axis line/.append style={black},every x tick label/.append style={font=\color{black}},every outer y axis line/.append style={black},every y tick label/.append style={font=\color{black}}},
			point meta min=0,
			point meta max=5
			]
			\addplot [forget plot] graphics [xmin=0.5,xmax=281.5,ymin=0.5,ymax=187.5] {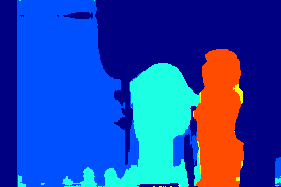};
			\end{axis}
		\end{tikzpicture}%
		}
		\subcaption{$\Delta x$ values from $E_{(-2,0)}$ and $E_{(2,0)}$ \label{fig:B4:4m:m2p2}}
		\vspace{.4cm}
	\end{minipage}
	\begin{minipage}[t]{\columnwidth}
		\centering
		\begin{tikzpicture}	
			\begin{axis}[%
			width=2.204861in,
			height=1.467292in,
			at={(0.853125in,0.823646in)},
			scale only axis,
			axis on top,
			separate axis lines,
			every outer x axis line/.append style={black},
			every x tick label/.append style={font=\color{black}},
			xmin=0.5,
			xmax=281.5,
			tick align=outside,
			every outer y axis line/.append style={black},
			every y tick label/.append style={font=\color{black}},
			y dir=reverse,
			ymin=0.5,
			ymax=187.5,
			hide axis,
			colormap/jet,
			colorbar,
			colorbar style={
				separate axis lines,every outer x axis line/.append style={black},every x tick label/.append style={font=\color{black}},every outer y axis line/.append style={black},every y tick label/.append style={font=\color{black}}},
			point meta min=0,
			point meta max=9
			]
			\addplot [forget plot] graphics [xmin=0.5,xmax=281.5,ymin=0.5,ymax=187.5] {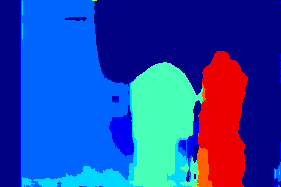};
			\end{axis}
		\end{tikzpicture}%
		\subcaption{$\Delta x$ values from $E_{(-4,0)}$ and $E_{(4,0)}$ \label{fig:B8:4m:m4p4}}
		\vspace{.2cm}
	\end{minipage}
	\begin{minipage}[t]{\columnwidth}
		\centering
		\begin{tikzpicture}
			\begin{axis}[%
			width=2.204861in,
			height=1.467292in,
			at={(0.853125in,0.823646in)},
			scale only axis,
			axis on top,
			separate axis lines,
			every outer x axis line/.append style={black},
			every x tick label/.append style={font=\color{black}},
			xmin=0.5,
			xmax=281.5,
			tick align=outside,
			every outer y axis line/.append style={black},
			every y tick label/.append style={font=\color{black}},
			y dir=reverse,
			ymin=0.5,
			ymax=187.5,
			hide axis,
			colormap/jet,
			colorbar,
			colorbar style={
				ytick={0,1,2,3,4,5},
				separate axis lines,every outer x axis line/.append style={black},every x tick label/.append style={font=\color{black}},every outer y axis line/.append style={black},every y tick label/.append style={font=\color{black}}},
			point meta min=0,
			point meta max=5
			]
			\addplot [forget plot] graphics [xmin=0.5,xmax=281.5,ymin=0.5,ymax=187.5] {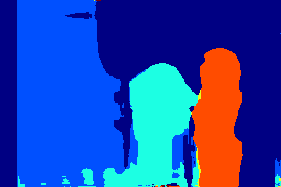};
			\end{axis}
		\end{tikzpicture}%
		\subcaption{$\Delta x$ values from $E_{(0,0)}$ and $E_{(4,0)}$ \label{fig:B4:4m:0p4}}
		\vspace{.2cm}
	\end{minipage}
	\caption[Disparity maps from sub-aperture images $E_{(i,g)}$ with $b_U > f_U$]{Disparity maps from sub-aperture images $E_{(i,g)}$ with $b_U > f_U$. \subref{fig:Base:4m} Central image $E_{(0,0)}$ containing $281$ by $187$ pixels; \subref{fig:B4:4m:m2p2} Disp. map with $G=4$, $\text{max}\{\Delta x\} = 5$ and block size = 33; \subref{fig:B8:4m:m4p4} Disp. map with $G=8$, $\text{max}\{\Delta x\} = 9$ and block size = 39; \subref{fig:B4:4m:0p4} Disp. map with $G=4$, $\text{max}\{\Delta x\} = 5$ and block size = 33. \label{fig:f197Base:4m}}
\end{figure}

Whenever $d_f \to \infty$, virtual camera tilt angles in our model are assumed to be $\Phi_G=0\degree$. Accurate baseline measurements inevitably confirm predicted tilt angles as measured baselines would deviate otherwise. To ensure this is the case, a second SPC triangulation experiment is carried out with $d_f \approx 4$~m, yielding images shown in Fig.~\ref{fig:f197Base:4m}.
%
\begin{table}[H]
	\small
	\centering
	\captionsetup{justification=centering}
	\caption[Tilt angle results $\Phi_G$ with 4~m focus $(b_U>f_U)$]{Tilt angle results $\Phi_G$ with 4~m focus ($b_U>f_U$).\label{tab:tiltComb}}
	\begin{minipage}[b]{.48\linewidth}
		\centering
		\renewcommand{\arraystretch}{1.2}
		\subcaption{$\Phi_4$ from \Cref{fig:B4:4m:m2p2,fig:B4:4m:0p4}\label{tab:tiltComb:a}}
		\resizebox{\linewidth}{!}{%
		\begin{tabular}{ c c c }
			\toprule
			\textbf{$\Delta x$} & \textbf{$Z_{G,\Delta x}$~[cm]} & \multicolumn{1}{c}{\textbf{Measured $\Phi_4$~[$\degree$]}} \\
			\midrule
			0 & 384 & 0.0429 \\
			1 & 218 & 0.0429 \\
			2 & 152 & 0.0429 \\
			4 & 95 & 0.0429 \\
			\bottomrule
		\end{tabular}
		}
	\end{minipage}
	\hfill
	\begin{minipage}[b]{.48\linewidth}
		\centering
		\renewcommand{\arraystretch}{1.2}
		\subcaption{$\Phi_8$ from Fig.~\ref{fig:B8:4m:m4p4}\label{tab:tiltComb:b}}
		\resizebox{\linewidth}{!}{%
		\begin{tabular}{ c c c }
			\toprule
			\textbf{$\Delta x$} & \textbf{$Z_{G,\Delta x}$~[cm]} & \multicolumn{1}{c}{\textbf{Measured $\Phi_8$ [$\degree$]}} \\
			\midrule
			0 & 384 & 0.0857 \\
			2 & 218 & 0.0857 \\
			4 & 152 & 0.0857 \\
			8 & 95 & 0.0857 \\
			\bottomrule
		\end{tabular}
		}
	\end{minipage}
	\\[4ex]
	\begin{minipage}[b]{\linewidth}
		\centering
		\renewcommand{\arraystretch}{1.2}
		\subcaption{Comparison of predicted and measured $\Phi_G$ where $d_f\approx 4~m$ \label{tab:tiltComb:e}}
		\resizebox{\linewidth}{!}{%
		\begin{tabular}{ c c c c c }
			\toprule
			& & \textbf{Predicted} & \textbf{Avg. measured} & \textbf{Deviation} \\
			& \textbf{$\Phi_G$} & \textbf{$\Phi_G$ [\degree]} & \multicolumn{1}{c}{\textbf{$\Phi_G$ [\degree]}} & \multicolumn{1}{c}{\textbf{$ERR_{\Phi_G}$~[\%]}} \\
			\midrule
			\multirow{2}{*}{\textbf{Proposed}} & $\Phi_4$ & 0.0429 & 0.0429 & 0.0000 \\
			& $\Phi_8$ & 0.0857 & 0.0857 & 0.0000 \\
			\midrule
			\multirow{2}{*}{\cite{Hahne:14:IEEE,Hahne:14:OPEX}} & $\Phi_4$ & 0.0429 & -0.3427 & 899.3410 \\
& $\Phi_8$ & 0.0857 & -0.6852 & 899.2393 \\
			\bottomrule
		\end{tabular}
		}
	\end{minipage}
	\captionsetup{justification=justified}
\end{table}
%
Disparity maps in \Cref{fig:B4:4m:m2p2,fig:B4:4m:0p4} give further indication that the spacing between adjacent virtual cameras is consistent. Results in Table~\ref{tab:tiltComb} demonstrate that tilt angle predictions match measurements. It is further shown that virtual cameras are rotated by small angles of less than a degree. Nevertheless, these tilt angles are non-negligible as they are large enough to shift the $\Delta x=0$ disparity plane from infinity to $d_f \approx 4$~m, which can be seen in Fig.~\ref{fig:f197Base:4m}. \par%
%
Generally, \Cref{tab:baseInfComb,tab:tiltComb} suggest that the adapted stereo triangulation concept proves to be viable in an SPC without measurable deviations if objects are placed at predicted distances. A maximum baseline is achieved with a short MLA focal length $f_s$, large micro lens pitch $p_M$, long main lens focal length $f_U$ and a sufficiently large entrance pupil diameter. \par%
%
%
A baseline approximation of the first-generation Lytro camera may be achieved with the aid of the metadata (*.json file) attached to each light field photograph as it contains information about the micro lens focal length $f_s=0.025$~mm, pixel pitch $p_p \approx 0.0014$~mm and micro lens pitch \linebreak$p_M \approx 0.0139$~mm, yielding $M=9.9286$ samples per micro image. The accommodated zoom lens provides a variable focal length $f_U$ in the range of 6.45~mm -- 51.4~mm (43~mm -- 341~mm as 35~mm-equivalent)~\citep{LYTRO14}. It is unclear whether the source refers to the main lens only or to the entire optical system including MLA. From this, hypothetical baseline estimates for the first-generation Lytro camera are calculated via \Cref{eq:BaselineSlope,eq:virtualCamera,eq:baseline} and given in Table~\ref{tab:lytro}. 
\begin{table}[H]
	\small
	\centering
	\caption[Baseline estimates of Lytro's $1^{\text{st}}$ generation camera]{Baseline estimates of Lytro's $1^{\text{st}}$ generation camera.}
	\label{tab:lytro}
	\renewcommand{\arraystretch}{1.2}
	\captionsetup{justification=centering}
	\begin{tabular}{cccc}
		\toprule
		\textbf{$f_s$~[mm]} & \textbf{$f_U$~[mm]} & \textbf{$B_1$~[mm]} & \textbf{$B_{8}$~[mm]}  \\
		\midrule
		0.025 & 6.45 & 0.3612 & 2.8896 \\
		0.025 & 51.4 & 2.8784 & 23.0272 \\
		\bottomrule
	\end{tabular}
\end{table}
Disparity analysis of perspective Lytro images should lead to baseline measures $B_G$ similar to those of the prediction. However, verification is impossible as the camera's automatic zoom lens settings (current principal planes and pupil locations) are undisclosed. Reliable measurements of such require disassembly of the main lens, which is impractical in the case of present-day Lytro cameras as main lenses are unmountable.%
\subsection{Simulation}
\label{ssec:sim}
To obtain quantitative measures, this section investigates the positioning of a virtual camera array by modelling a plenoptic camera in an optics simulation software~\citep{Zemax}. Table~\ref{tab:baselineSim} reveals a comparison of predicted and simulated virtual camera positions just as their baseline $B_{G}$ and relative tilt angle $\Phi_G$. Thereby, the distance from an objective's front vertex $V_{1U}$ to entrance pupil $A''$ is given by
\begin{align}
	\overline{V_{1U}A''} = \overline{V_{1U}H_{1U}} + \overline{A''H_{1U}} 
\end{align}
bearing in mind that $\overline{A''H_{1U}}$ is the distance from entrance pupil $A''$ to object-side principal plane $H_{1U}$ and $\overline{V_{1U}H_{1U}}$ separates $H_{1U}$ from the front vertex $V_{1U}$. Simulated $\overline{V_{1U}A''}$ are obtained by extending ray slopes $q_{i,j}$ towards the sensor whilst these virtually elongated rays are seen to ignore lenses and finding the intersection of $q_{i,j}$ and $q_{i,j+1}$.
%
\begin{table*}[ht]
	\centering
	\caption[Baseline and tilt angle simulation with $G=6$ and $i=0$]{Baseline and tilt angle simulation with $G=6$ and $i=0$.}
	\label{tab:baselineSim}
	\renewcommand{\arraystretch}{1.5}
	\resizebox{\linewidth}{!}{%
	\begin{tabular}{ 
		|c|c|c
		|S[table-format=3.4]
		|S[table-format=2.4]
		|S[table-format=-1.4]
		|S[table-format=3.4]
		|S[table-format=2.4]
		|S[table-format=-1.4]
		|S[table-format=1.4]
		|S[table-format=1.4]
		|S[table-format=1.4]|
	}
		\hline
		\multicolumn{3}{|c|}{\textbf{Setup}} &
		\multicolumn{3}{c|}{\textbf{Prediction}} & \multicolumn{3}{c|}{\textbf{Simulation}} &
		\multicolumn{3}{c|}{\textbf{Deviation} [\%]} \\ 
		\hline
		{$d_f$} & {$f_U$} & {$f_s$} & {$\overline{V_{1U}A''}$ [mm]} & {$B_{G}$ [mm]} & {$\Phi_i$ [$^{\circ}$]} & {$\overline{V_{1U}A''}$ [mm]} & {$B_{G}$ [mm]} & {$\Phi_i$ [$^{\circ}$]} & {$ERR_{\overline{V_{1U}A''}}$} & {$ERR_{B_{G}}$} & {$ERR_{\Phi_i}$} \\ 
		\hline
		\multirow{3}{*}{\begin{turn}{90}Inf\end{turn}}
		& $f_{193}$ & {(\RM{2})} & 240.2113 & 3.7956 & 0.0000 & 240.1483 & 3.7949 & 0.0000 & 0.0262 & 0.0184 & {--} \\ 
		& $f_{90}$ & {(\RM{2})} & 27.4627 & 1.7752 & 0.0000 & 27.4081 & 1.7748 & 0.0001 & 0.1988 & 0.0225 & {--} \\ 
		& $f_{193}$ & {(\RM{1})} & 240.2113 & 8.3503 & 0.0000 & 239.3988 & 8.3450 & 0.0000 & 0.3382 & 0.0635 & {--} \\ 
		\hline
		\multirow{3}{*}{\begin{turn}{90}3 m\end{turn}}
		& $f_{193}$ & {(\RM{2})} & 240.2113 & 4.2748 & -0.0816 & 239.8612 & 4.2738 & -0.0816 & 0.1457 & 0.0234 & 0.0000 \\ 
		& $f_{90}$ & {(\RM{2})} & 27.4627 & 1.8357 & -0.0361 & 27.3309 & 1.8352 & -0.0360 & 0.4799 & 0.0272 & 0.2770 \\ 
		& $f_{193}$ & {(\RM{1})} & 240.2113 & 9.4047 & -0.1795 & 238.9043 & 9.3964 & -0.1795 & 0.5441 & 0.0883 & 0.0000 \\ 
		\hline
		\multirow{3}{*}{\begin{turn}{90}1.5 m\end{turn}}
		& $f_{193}$ & {(\RM{2})} & 240.2113 & 4.9097 & -0.1897 & 239.6932 & 4.9078 & -0.1897 & 0.2157 & 0.0387 & 0.0000 \\ 
		& $f_{90}$ & {(\RM{2})} & 27.4627 & 1.9049 & -0.0774 & 27.2150 & 1.9042 & -0.0773 & 0.9020 & 0.0367 & 0.1292 \\ 
		& $f_{193}$ & {(\RM{1})} & 240.2113 & 10.8014 & -0.4173 & 238.1212 & 10.7866 & -0.4173 & 0.8701 & 0.1370 & 0.0000 \\ 
		\hline
	\end{tabular}
	}
\end{table*}
Observations in Table~\ref{tab:baselineSim} indicate that the baseline grows with
\begin{itemize}
	\item larger main lens focal length $f_U$
	\item shorter micro lens focal length $f_s$
	\item decreasing focusing distance $d_f$ $(a_U)$
\end{itemize}
given that the entrance pupil diameter is large enough to accommodate the baseline. Besides, it has been proven that tilt angle rotations become larger with decreasing $d_f$. Baselines have been estimated accurately with errors below $0.1~\%$ on average except for one example. The key problem causing the largest error is that MLA~(\RM{1}) features a shorter focal length $f_s$ than MLA~(\RM{2}) which produces steeper light ray slopes $m_{c+i,j}$ and hence severe aberration effects. Tilt angle errors remain below $0.3~\%$ although results deviate by only $0.001\degree$ for $f_{90}$ and are even non-existent for $f_{193}$. However, entrance pupil location errors of about $\leq 1~\%$ are larger than in any other simulated validation. One reason for these inaccuracies is that the entrance pupil $A''$ is an imaginary vertical plane which in reality may exhibit a non-linear shape around the optical axis.\par %
%
\begin{table*}[ht]
	\centering
	\caption[Disparity and distance simulation with $G=1$ and $i=0$]{Disparity simulation and distance with $G=1$ and $i=0$.}
	\label{tab:disparitySim}
	\renewcommand{\arraystretch}{1.5}
	\resizebox{\linewidth}{!}{%
	\begin{tabular}{ 
		      |c|c
		      |S[table-format=5.4]
		      |S[table-format=4.4]
		      |S[table-format=4.4]
		      |S[table-format=5.4]
		      |S[table-format=4.4]
		      |S[table-format=4.4]
		      |S[table-format=-1.4]
		      |S[table-format=-1.4]
		      |S[table-format=1.4]|
		}
		\hline
		\multicolumn{2}{|c|}{\textbf{Setup}} &
		\multicolumn{3}{c|}{\textbf{Prediction}} & \multicolumn{3}{c|}{\textbf{Simulation}} &
		\multicolumn{3}{c|}{\textbf{Deviation}} \\ 
		\hline
		\multirow{2}{*}{$d_f$} & \multirow{2}{*}{$\Delta x$} & \multicolumn{3}{c|}{$Z_{1,\Delta x}$ [mm]} & \multicolumn{3}{c|}{$Z_{1,\Delta x}$ [mm]} & \multicolumn{3}{c|}{$ERR_{Z_{1,\Delta x}}$ [\%]} \\ 
		\cline{3-11}
		& & $f_{193}$ { \& (\RM{2})} & $f_{90}$ { \& (\RM{2})} & $f_{193}$ { \& (\RM{1})} & $f_{193}$ { \& (\RM{2})} & $f_{90}$ { \& (\RM{2})} & $f_{193}$ { \& (\RM{1})} & $f_{193}$ { \& (\RM{2})} & $f_{90}$ { \& (\RM{2})} & $f_{193}$ { \& (\RM{1})} \\
		\hline
		\multirow{3}{*}{\begin{turn}{90}Inf\end{turn}}
		& 0 & {$\infty$} & {$\infty$} & {$\infty$} & {$\infty$} & {$\infty$} & {$\infty$} & {--} & {--} & {--} \\ 
		& 1 & 978.2150 & 213.9790 & 2152.0729 & 978.2797 & 213.9573 & 2151.2840 & -0.0066 & 0.0101 & 0.0367 \\ 
		& 2 & 489.1075 & 106.9895 & 1076.0365 & 489.1026 & 106.9431 & 1075.1177 & 0.0010 & 0.0434 & 0.0854 \\ 
		\hline
		\multirow{3}{*}{\begin{turn}{90}3~m\end{turn}}
		& 0 & 3001.4530 & 2913.5460 & 3001.4530 & 3000.8133 & 2923.2193 & 2999.3120 & 0.0213 & -0.3320 & 0.0713 \\ 
		& 1 & 877.9068 & 212.1505 & 1429.6116 & 877.4653 & 212.0285 & 1427.8084 & 0.0503 & 0.0575 & 0.1261 \\ 
		& 2 & 514.1456 & 110.0831 & 938.2541 & 513.8952 & 109.9610 & 937.1572 & 0.0487 & 0.1109 & 0.1169 \\ 
		\hline
		\multirow{4}{*}{\begin{turn}{90}1.5~m\end{turn}}
		& -1 & 15770.8729 & {--} & 2521.0686 & 15764.1482 & {--} & 2517.6509 & 0.0426 & {--} & 0.1356 \\ 
		& 0 & 1482.8768 & 1410.2257 & 1482.8768 & 1482.3969 & 1412.2221 & 1481.1620 & 0.0324 & -0.1416 & 0.1156 \\ 
		& 1 & 778.0154 & 209.7424 & 1050.3402 & 777.8168 & 209.5320 & 1049.3327 & 0.0255 & 0.1003 & 0.0959 \\ 
		& 2 & 527.3487 & 113.2965 & 813.1535 & 527.0279 & 113.0602 & 811.8298 & 0.0608 & 0.2086 & 0.1628 \\ 
		\hline
	\end{tabular}
	}
\end{table*}
An experiment assessing the relationship between disparity $\Delta x$ and distance $Z_{G,\Delta x}$ using different objective lenses is presented in Table~\ref{tab:disparitySim}. From this, it can be concluded that denser depth sampling is achieved with larger main lens focal length $f_U$. Moreover, it is seen that a tilt in virtual cameras yields a negative disparity $\Delta x$ for objects further away than $d_f$ which is a phenomenon that also applies to tilted cameras in stereoscopy. %
The reason why $d_f \approx Z_{G,\Delta x}$ when $\Delta x=0$ is that $Z_{G,\Delta x}$ reflects the separation between ray intersection and entrance pupil $A''$, which lies nearby the sensor and $d_f$ is the spacing between ray intersection and MLA's front vertex. 
Overall, it can be stated that distance estimates based on the stereo triangulation behave similar to those in geometrical optics with errors of up to $\pm0.33~\%$. %
%
\section{Discussion and Conclusions}
\label{sec:4}
%
In essence, this paper presented the first systematic study on how to successfully apply the triangulation concept to a Standard Plenoptic Camera (SPC). It has been shown that an SPC projects an array of virtual cameras along its entrance pupil, which can be seen as an equivalent to a multi-view camera system. Thereby, the proposed geometry of the SPC's light field suggests that the entrance pupil diameter constrains the maximum baseline. This backs up and further refines an observation made by \citeauthor*{AW}~(\citeyear{AW}), who considered the aperture size to be the baseline limit. Our customised SPC merely offers baselines in the millimetre range, which results in relatively small stereo vision setups. Due to this, depth sampling planes move towards the camera, which will prove to be useful for close range applications such as microscopy. It is also expected that multiple viewpoints taken with small baselines evade the occlusion problem. %
\par
The presented work has provided the first experimental baseline and distance results based on disparity maps obtained by a plenoptic camera. Predictions of our geometrical model match measures of the experimentation without indicating a significant deviation. An additional benchmark test of the proposed model with an optical simulation software has revealed errors of up to $\pm0.33~\%$ for baseline and distance estimates under different lens settings, which supports the model's accuracy. Deviations are due to the imperfections of objective lenses. More specifically, prediction inaccuracies may be caused by all sorts of aberrations that result in a non-geometrical behaviour of a lens. 
By compensating for this through enhanced image calibration, we believe it is possible to lower the measured deviation.
\par
\par
The major contribution of the proposed ray model is that it allows any SPC to be used as an object distance estimator. A broad range of applications for which stereoscopy has been traditionally occupied can benefit from this solution. This includes endoscopes or microscopes that require very close depth ranges, the automotive industry where tracking objects in road traffic is a key task and the robotics industry with robots in space or automatic vacuum cleaners at home. Besides this, plenoptic triangulation may be used for quality assurance purposes in the large field of machine vision. %
The model further assists in the prototyping stage of plenoptic photo and video cameras as it allows the baseline to be adjusted as desired. %
\par
Further research may investigate how triangulation applies to other types of plenoptic cameras, such as the focused plenoptic camera or coded-aperture camera. More broadly, research is also required to benchmark a typical plenoptic camera's depth resolution with that of competitive depth sensing techniques like stereoscopy, time of flight and light sectioning. %
\bibliographystyle{spbasic}      
\bibliography{template}   

\end{document}